\documentclass[twocolumn,superscriptaddress, amsmath,amssymb,aps,prb,floatfix]{revtex4-2}
\usepackage[usenames,dvipsnames]{color}

\usepackage{comment}
\usepackage{cancel}
\usepackage{ulem}
\usepackage{physics}
\usepackage{graphicx}% Include figure files
\usepackage{dcolumn}% Align table columns on decimal point
\usepackage{bm}% bold math
\usepackage{gensymb}
\usepackage{tensor}
\usepackage{rotating}
\usepackage{multibib}
\usepackage[percent]{overpic}
\usepackage[breaklinks=true,colorlinks,citecolor=blue,linkcolor=blue,urlcolor=blue]{hyperref}% add hypertext capabilities
\begin{document}
\title{Recognizing molecular chirality via twisted 2D materials}
\date{\today}
\author{Lorenzo Cavicchi}
\affiliation{Scuola Normale Superiore, I-56126 Pisa,~Italy}
\author{Mayra Peralta}
\affiliation{Max Planck Institute for Chemical Physics of Solids, N\"{o}thnitzer Str. 40, Dresden 01187,~Germany}
\author{\'{A}lvaro Moreno}
\affiliation{ICFO-Institut de Ci\`{e}ncies Fot\`{o}niques, The Barcelona Institute of Science and Technology, Av. Carl Friedrich Gauss 3, 08860 Castelldefels (Barcelona),~Spain}
\author{Maia Vergniory}
\affiliation{Donostia International Physics Center, 20018 Donostia-San Sebastian,~Spain}
\author{Pablo Jarillo-Herrero}
\affiliation{Department of Physics, Massachusetts Institute of Technology, Cambridge, Massachusetts 02139,~USA}
\author{Claudia Felser}
\affiliation{Max Planck Institute for Chemical Physics of Solids, N\"{o}thnitzer Str. 40, Dresden 01187,~Germany}
\author{Giuseppe C. La Rocca}
\affiliation{NEST, Scuola Normale Superiore, I-56126 Pisa,~Italy}
\author{Frank H. L. Koppens}
\affiliation{ICFO-Institut de Ci\`{e}ncies Fot\`{o}niques, The Barcelona Institute of Science and Technology, Av. Carl Friedrich Gauss 3, 08860 Castelldefels (Barcelona),~Spain}
\affiliation{ICREA-Instituci\'{o} Catalana de Recerca i Estudis Avan\c{c}ats, Passeig de Llu\'{i}s Companys 23, 08010 Barcelona,~Spain}
\author{Marco Polini}
\affiliation{Dipartimento di Fisica dell'Universit\`a di Pisa, Largo Bruno Pontecorvo 3, I-56127 Pisa, Italy}
\affiliation{ICFO-Institut de Ci\`{e}ncies Fot\`{o}niques, The Barcelona Institute of Science and Technology, Av. Carl Friedrich Gauss 3, 08860 Castelldefels (Barcelona),~Spain}
\begin{abstract}
Chirality pervades natural processes from the atomic to the cosmic scales, crucially impacting molecular chemistry and pharmaceutics. Traditional chirality sensing methods face challenges in sensitivity and efficiency, prompting the quest of novel chiral recognition solutions based on nanophotonics. In this work we theoretically investigate the possibility to carry out  enantiomeric discrimination by measuring the spontaneous emission rate of chiral molecules on twisted two-dimensional materials. We first present a general theoretical framework based on dyadic Green's functions to calculate the chiral contribution to the decay rate in the presence of a generic chiral bilayer interface. We then combine this theory with density functional theory to obtain numerical estimates of the decay rate of helical bilayer nanographene molecules placed on top of twisted bilayer graphene.
\end{abstract}
\maketitle
\section{Introduction}
\label{sect:intro}

Chirality, the inherent asymmetry of molecules, plays a role in various natural processes, from microscopic molecules like amino acids and DNA—the building blocks of life—to plants, animals, and even macroscopic celestial objects like galaxies. In fields like chemistry and pharmaceutics, where identifying the chirality of molecules plays a crucial role~\cite{hutt_1996}, traditional methods for chirality sensing, such as circular dichroism (CD) spectroscopy~\cite{price_2005} and methods based on optical rotatory dispersion~\cite{yoo_2019}, often face limitations in sensitivity and efficiency. Recent advances in nanophotonics offer promising solutions for overcoming these challenges by exploiting the interaction between light and matter at the nanoscale~\cite{lininger_2023}. For example, methodologies based on superchiral near-fields~\cite{hendry_2010}, CD enhancement via twisted optical metamaterials~\cite{zhao_2017} or plasmons~\cite{wang_2022}, coupling to chiral cavities~\cite{mauro_2023, riso_2023} and self-assembled metal nanoparticles~\cite{wang_2024} have been proposed as avenues for chiral sensing. The sensitivity limitation that these techniques often face~\cite{barron_book}, makes chirality detection a big technological challenge.

In this Article, we explore the possibility of enantiomeric discrimination based on the modification of the spontaneous emission (SE) of a chiral quantum emitter with chiral bilayer substrates. Moir\'{e} materials obtained by twisting atomically thin crystals with respect to each other~\cite{Andrei2020,Andrei2021,Kennes2021} have attracted a great deal of interest since the discovery of superconductivity~\cite{Cao2018a} and correlated insulating states~\cite{Cao2018b} in twisted bilayer graphene (TBG)~\cite{Santos2007,shallcross_prl_2008,mele_prb_2010,Li2010,shallcross_prb_2010,morell_PRB_2010,bistritzer_prb_2010,Bistritzer2011,lopes_prb_2012}, and regimes of fractional quantum anomalous Hall in twisted bilayer MoTe$_2$~\cite{Cai2023,Park2023,Zeng2023,Xu2023} and superconductivity~\cite{xia_arxiv_2024, guo_arxiv_2024} in twisted bilayer WSe$_2$. Despite these groundbreaking discoveries, very few studies have highlighted the fact that these systems are {\it chiral materials}. 
Natural optical activity~\cite{Landau08}, a hallmark of chiral materials, was, in fact, experimentally discovered in TBG~\cite{Kim2016}. TBG exhibits a remarkably large circular dichroism~\cite{Kim2016,Morell_2DMaterials_2017}, up to a factor $100$ stronger than for a layer of chiral molecules of similar thickness. This effect harbors from the peculiar transport properties of twisted bilayers: an inter-layer Hall conductivity appears due to chirality~\cite{Morell_2DMaterials_2017,stauber_2018,ho_2023,ding_2023}. Near-field plasmonic superchiral electromagnetic fields~\cite{tang_prl_2010} have also been predicted in TBG~\cite{stauber_nanolett_2020}. These fields can be used, in principle, to exploit the light absorption dissymmetry between adsorbed left and right chiral molecules on the chiral substrate itself~\cite{hendry_naturenanotech_2010}. 

Inspired by Curie's dyssymetry principle, stated by Pierre Curie at the end of the XIX century~\cite{curie_1894}, we investigate the modification of the SE of a chiral molecule due to the proximity to a chiral bilayer. It was Purcell~\cite{purcell_1946} the first to understand that the surrounding environment has the potential to significantly alter the SE rate of a quantum emitter. Since then, the study of this effect, which was named the Purcell effect, has been an active field of research~\cite{lin_1992,klimov_1999,tomas_2001,blanco_2004,carminati_2006,rosa_2008,biehs_2011,kort-kamp_2013,park_2019,farina_2019,abrantes_2023,koppens_nano_letters_2011, kort-kamp_prb_2015}. Here we combine Curie's principle to the Purcell effect, proposing chiral bilayers as a suitable electromagnetic environment to exploit the chirality of a quantum emitter through the asymmetric modification of its fluorescence properties. The concept of using twisted bilayer substrates as a local chemical environment to enhance chemical reactivity is also being recently explored~\cite{zhang_PRB_2024}.

%In particular, graphene substrates have been shown to be excellent platforms for several nanophotonic applications. In fact, graphene surface plasmons can not only modify the SE rate~\cite{koppens_nano_letters_2011, kort-kamp_prb_2015}, but also enhance the electromagnetic field emitted by quantum emitters~\cite{hanson_2012}, mediate superradiance between two quantum emitters~\cite{huidobro_2012}, and affect energy transfer between quantum systems~\cite{karanikolas_2015}. 
%
%In their seminal work~\cite{tang_prl_2010}, Tang e Cohen, showed how super-chiral electromagnetic fields can be obtained in the proximity of planar chiral materials. 

This Article is organized as follows: Section~\ref{sectionII} collects the main theoretical results for the SE of a chiral molecule in a generic inhomogeneous environment; Section~\ref{sectionIII} resumes the minimal description of the optical conductivity of a chiral (homo)bilayer and the derivation of its reflection properties as a substrate; finally, Section~\ref{sectionIV} presents the main numerical results. The main text is accompanied with a substantial set of appendices where the reader can find additional details on the theoretical framework and calculations, as well as further numerical results.

\begin{figure}[t]
    \centering
    \begin{overpic}[width=0.5\textwidth]{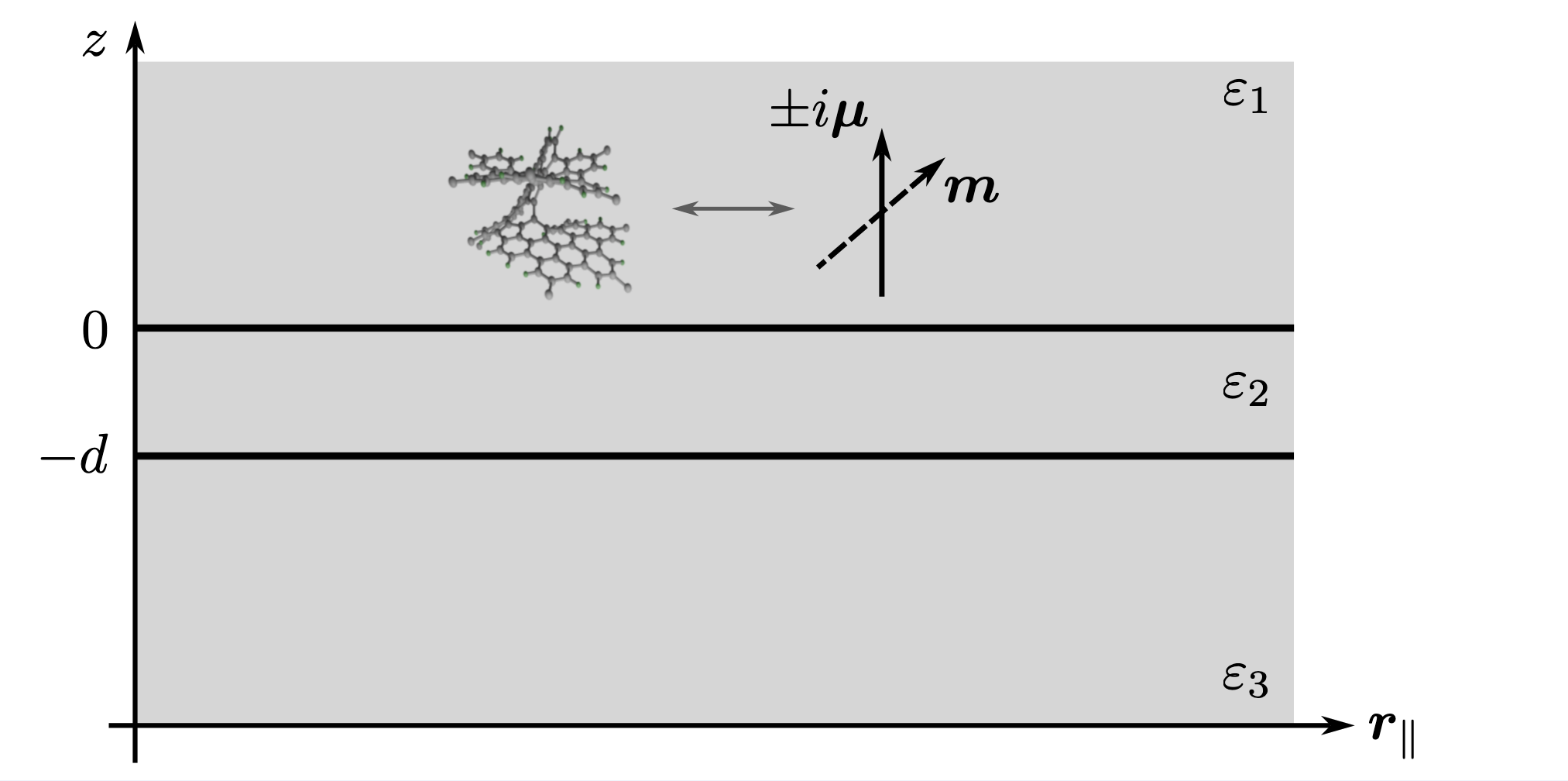}
    \put(84.9,25.5){Chiral}
    \put(85,21.5){bilayer}
    \end{overpic}
    \caption{The geometry considered in this work. The chiral molecule is placed in an inhomogeneous environment composed of two dielectric semi-spaces, with dielectric constants $\varepsilon_1$ (top) and $\varepsilon_3$ (bottom), and magnetic permeability $\mu=1$. These two regions are separated by a conducting and translationally invariant chiral bilayer interface, macroscopically described by a conductivity tensor $\hat{\bm \sigma}(\omega)$. The top layer is placed at $z=0$ and the bottom one at $z = -d$. The dielectric constant in the space $-d<z<0$ is fixed to $\varepsilon_2$ ($\mu=1$). We fix the molecule height as $z=z_0$. The spectral properties of the molecules can be described by a superposition of an electric $\bm \mu$ and a magnetic $\bm{m}$ dipole.}
    \label{fig1}
\end{figure}
\section{Decay rate of a chiral molecule in inhomogeneous environments}\label{sectionII}
Consider a quantum emitter placed in an electromagnetic environment composed by three regions of space, as shown in Fig.~\ref{fig1}. Within the dipole approximation~\cite{novotny}, the optical transition of a chiral molecule can be described as a superposition of an electric and magnetic dipole, that is, a chiral dipole $\bm{d}_{\pm}(\bm{r}, t) = (\pm i \bm{\mu} + \bm{m})\delta(\bm{r} -\bm{r}_0)e^{-i\omega t}$~\cite{meskers_2022,eismann_optica_2018}, where $\hbar\omega$ is the energy of the transition. The ``$\pm$'' sign represents the chirality (or handness) of the molecule, corresponding to right- or left-handed enantiomers. The rate of SE for such a molecule is influenced by its local environment. The SE rate $\Gamma$ is related to the power dissipated by the classical dipole, denoted as $P$, according to the relation~\cite{novotny}:
\begin{equation}\label{eqn:Gamma_P}
    \frac{\Gamma}{\Gamma_0} = \frac{P}{P_0}~.
\end{equation}
Eq.~\eqref{eqn:Gamma_P} relates the quantum-mechanical picture of SE (the left-hand side) to the classical formalism of dipole radiation (the right-hand side). Thanks to this relationship we can calculate the relative change of the spontaneous decay rate of a quantum emitter in an inhomogeneous environment within a classical approach~\cite{novotny,koppens_nano_letters_2011, kort-kamp_prb_2015}. The dissipated power of a chiral dipole $\bm{d}_{\pm}(\bm{r}, t)$ can be subdivided into four contributions: an electric-electric, magnetic-magnetic, electro-magnetic and mangeto-electric term,~i.e. $P = P_{e,e} + P_{m,m} + P_{e,m} + P_{m,e}$. One can manipulate the dissipated power in order to extract the contributions coming from the electromagnetic environment~\cite{novotny} (see also Appendix~\ref{appendixA}):
\begin{equation}\label{eqn:total_P}
    P = P_0 + P_{e,e}^{\rm ref} + P_{m,m}^{\rm ref} + P_{e,m}^{\rm ref} + P_{m,e}^{\rm ref}~,
\end{equation}
where $P_0$ is the ``free-space'' contribution, given by~\cite{novotny} 
\begin{equation}
    P_0(\omega) = \frac{\omega^4}{3c^3}\left(\frac{|\bm{\mu}|^2}{\varepsilon_1} + |\bm{m}|^2\right)~,
\end{equation}
while $P_{e,e}^{\rm ref}$, $P_{m,m}^{\rm ref}$, $P_{e,m}^{\rm ref} $ and $ P_{m,e}^{\rm ref}$ describe the contributions coming from the reflected electromagnetic (EM) field from the substrate, and they are a function of the fluorescence frequency $\omega$ and the distance between the molecule and the interface $z$. A straightforward calculation (Appendix~\ref{appendixA}) that relies on the Dyadic Green functions (DGfs) approach to electromagnetism~\cite{novotny, Tai1994, peres}, leads to the following explicit expressions for the environment contributions to the total dissipated power (sum over repeated indices is intended):
\begin{widetext}
\begin{gather}
    P_{e,e}^{\rm ref}(z_0,\omega) = \frac{\omega^3}{2c^2}\int\frac{d^2\bm{k}}{(2\pi)^2}\Im \left[\mu_i  G_{{\rm ref},ij}(\bm{k},z_0,z_0,\omega) \mu_j\right]~,\label{eqn:ref_power1} \\
    P_{m,m}^{\rm ref}(z_0,\omega) = \frac{\omega^3}{2c^3}\int\frac{d^2\bm{k}}{(2\pi)^2}\Im \left[m_i G_{{\rm ref},ij}^{\rm M}(\bm{k},z_0,z_0,\omega) m_{j}\right]\label{eqn:ref_power4}~,\\
    P_{e,m}^{\rm ref}(z_0,\omega) = \pm \frac{\omega^2}{2c}\int\frac{d^2\bm{k}}{(2\pi)^2}\Im \left[\mu_i \epsilon_{ijk}\left.\partial_{j} G_{{\rm ref},k\ell}^{\rm M}(\bm{k},z_0,z_0,\omega)\right|_{z=z_0} m_{\ell}\right]~,\label{eqn:ref_power2} \\
    P_{m,e}^{\rm ref}(z_0,\omega) = \pm\frac{\omega^2}{2c}\int\frac{d^2\bm{k}}{(2\pi)^2}\Im \left[m_i\epsilon_{ijk}\left.\partial_j G_{{\rm ref},k\ell}(\bm{k},z_0,z_0,\omega) \right|_{z=z_0}\mu_\ell\right]\label{eqn:ref_power3}~,
\end{gather}
\end{widetext}
where the electric $G_{{\rm ref},ij}(\bm{r},\bm{r}^\prime,\omega)$ and magnetic $G_{{\rm ref},ij}^{\rm M}(\bm{r},\bm{r}^\prime,\omega)$ DGfs are the physical response functions that allow one to express the electric $\bm{E}_{{\rm ref}}({\bm r}, \omega)$ and magnetic $\bm{H}_{{\rm ref}}({\bm r}, \omega)$ fields, that are {\it reflected} by the substrate, induced by an external current ${\bm J}({\bm r}^\prime,\omega)$ located at a point ${\bm r}^\prime$ at any point ${\bm r}$ in space. (We refer the reader to the Appendix~\ref{appendixB} for further details on this formalism). Here, since the system has cylindrical symmetry due to the translational invariance along the interface plane, see also Fig.~\ref{fig1}, we introduced the Fourier transform of the electric and magnetic DGfs:
\begin{equation}
\begin{split}
    G_{{\rm ref},ij}^{({\rm M})}({\bm r}_\parallel,&{\bm r}_\parallel^\prime,z,z^\prime,\omega) =\\
    &\int\frac{d\bm k}{(2\pi)^2} e^{i\bm{k}\cdot(\bm{r}_\parallel-\bm{r}_\parallel^\prime)} G_{{\rm ref},ij}^{({\rm M})}(\bm k,z,z^\prime,\omega)~.
\end{split}
\end{equation}
Although contributions~\eqref{eqn:ref_power1} and~\eqref{eqn:ref_power4} to the total dissipated power have already been discussed in the literature~\cite{koppens_nano_letters_2011,kort-kamp_prb_2015,novotny,wiecha_prb_2018}, the cross contributions, Eq.~\eqref{eqn:ref_power2} and Eq.~\eqref{eqn:ref_power3}, to the best of our knowledge, have been previously mentioned in the context of tailoring the SE of chiral molecules in chiral cavities~\cite{voronin_ACSPHOTONICS_2022} but never quantitatively discussed~\cite{chiral_purcell_footnote}. Crucially, these two contributions contain the chiral information of the emitting source $\bm{d}_{\pm}$ and, as we shall discuss momentarily, also depend on the chirality of the substrate. 

The full expressions of the reflected electric $G_{{\rm ref},ij}(\bm{k},z,z^\prime,\omega)$ and magnetic $G_{{\rm ref},ij}^{\rm M}(\bm{k},z,z^\prime,\omega)$ DGf are derived in Appendix~\ref{appendixC}.
\subsection{Vertical dipole}
As a case of study, we consider a chiral dipole $\bm{d}_{\pm}$ that consists of a vertical electric dipole and an inclined magnetic moment: $\bm{d}_{\pm} = \left\{(\pm i)\mu_0\hat{\bm z} + m_0[\hat{\bm x}\sin\varphi + \hat{\bm z}\cos\varphi]\right\}e^{-i\omega t}$. The perfect alignment is obtained when $\varphi = 0$. The expressions for the substrate contributions~\eqref{eqn:ref_power1}-\eqref{eqn:ref_power3} to the dissipated power are obtained by exploiting the reflection part of the DGfs (see Appendix~\ref{appendixC}). The electric-electric and magnetic-magnetic contributions are:
\begin{gather}
    \begin{split}
    \label{eqn:Pee}
    P_{e,e}^{\rm ref}(z_0,\omega) = \frac{\omega^3}{2 c^2}\mu_0^2 \int&\frac{d^2\bm{k}}{(2\pi)^2}\times\\
    &\Im \left[\frac{2\pi i k^2}{k_1^2k_{z,1}} e^{2ik_{z,1}z_0}r^{p,p}\right]
    \end{split}~,\\
    \begin{split}
    \label{eqn:Pmm}
    &P_{m,m}^{\rm ref}(z_0,\omega) =\frac{\omega^3}{2 c^2}m_0^2\int\frac{d^2\bm{k}}{(2\pi)^2}\times\\
    &\Im \left[\frac{2\pi i}{k_1^2k_{z,1}} e^{2ik_{z,1} z_0}r^{s,s}(k^2 \cos^2\varphi-k_{z,1}^2 \sin^2\varphi)\right]~,
    \end{split}
\end{gather}
where $k_n = \sqrt{\varepsilon_n}\omega/c$ ($n=1,2,3$ here refers to the region of space, see Fig.~\ref{fig1}) is the wavevector in the region $z>0$, the transverse wave-vector is defined by:
\begin{equation}\label{eqn:transverse_wavevector}
k_{z,n} =
\begin{cases}
    \sqrt{k_n^2 - k^2}, \qquad {\rm if} \quad k<k_n~,\\
    i\sqrt{k^2-k_n^2}, \qquad {\rm if} \quad k>k_n~.
\end{cases}
\end{equation}
In Eqs.~\eqref{eqn:Pee} and~\eqref{eqn:Pmm} we introduced the tensor $r^{\alpha,\beta}$ that describes the reflection at the interface. It is defined as the tensor whose elements are the amplitudes of the reflected electric field with $\alpha-$ polarization when a unitary amplitude field with $\beta-$ polarization impinges on the interface. Here $\alpha,\beta = s,p$ refer to the $s-$ and $p-$ polarizations,~i.e. those associated to transverse and parallel modes with respect to the plane of incidence, respectively (see also Appendix~\ref{appendixC}). For $\varphi = 0$, these expressions reproduce the well-known results for the vertical dipole~\cite{novotny,koppens_nano_letters_2011}.

Similarly to $P_{e,e}^{\rm ref}$ and $P_{m,m}^{\rm ref}$, we derive an expression for the cross terms $P_{e,m}^{\rm ref}$ and $P_{m,e}^{\rm ref}$:
\begin{widetext}
\begin{gather}
    \label{eqn:Pem}
    P_{e,m}^{\rm ref}(z_0,\omega) = \pm \frac{\omega^2}{4c}\mu_0 m_0\int\frac{d^2\bm{k}}{(2\pi)^2} \Im\left[\frac{4\pi k}{k_1 k_{z,1}}e^{2 i k_{z,1} z_0} r^{p,s} (k \cos \varphi +k_{z,1} \sin\varphi)\right]~.\\
    \label{eqn:Pme}
    P_{m,e}^{\rm ref}(z_0,\omega) = \mp \frac{\omega^2}{4c}\mu_0 m_0\int\frac{d^2\bm{k}}{(2\pi)^2} \Im\left[\frac{4\pi  k}{k_1 k_{z,1}}e^{2 i k_{z,1} z_0} r^{s,p} (k \cos \varphi +k_{z,1} \sin\varphi)\right]~.
\end{gather}
\end{widetext}
Eqs.~\eqref{eqn:Pem} and~\eqref{eqn:Pme} describe how the chirality of the substrate modifies the dissipated power of the chiral emitter and hence its decay rate. Since the total dissipated power is the sum of all contributions~\eqref{eqn:Pee},~\eqref{eqn:Pmm},~\eqref{eqn:Pem} and~\eqref{eqn:Pme}, a substrate whose reflection tensor $r^{\alpha,\beta}$ is diagonal,~i.e. $r^{p,s} = r^{s,p} = 0$, or symmetric,~i.e. $r^{p,s} = r^{s,p}$, makes the decay rate {\it not sensitive} to the chirality of the molecule itself. A chiral substrate shows an anti-symmetric reflection tensor~\cite{ali_jopt_1992,li_2000,buhmann_IJMPA_2016},~that is, $r^{p,s} = -r^{s,p}$ (Appendix~\ref{appendixC1}). So, in this case, the expressions~\eqref{eqn:Pem} and~\eqref{eqn:Pme} sum up giving a finite contribution to the dissipated power. We also notice that, when the handness of the chiral substrate changes, the cross reflection terms $r^{p,s}$ and $r^{s,p}$ change sign, showing that both $P_{e,m}$ and $P_{m,e}$ are {\it sensitive} to the chirality of the substrate itself.

For results concerning an arbitrarily oriented dipole and further details on the derivation of the results presented in this section, we refer the reader to the Appendix~\ref{appendixD}.
\section{Chiral response of chiral bilayers}\label{sectionIII}
A twisted homobilayer is a van der Waals structure obtained by stacking two layers of the same kind and giving them a relative twist angle $\theta$. This stacking procedure can produce either a right-handed or left-handed bilayer depending on the direction of the twist, counterclockwise ($\theta$) or clockwise ($-\theta$). In this Section we first introduce the conductivity tensor of a twisted homobilayer. Then we use it to solve the scattering problem of an incident EM field by deriving the reflection tensor $r^{\alpha,\beta}$.
\subsection{Optical conductivity}
The optical conductivity of a 2D system is defined as the tensor relating the electric current $\bm{J}(\bm{r}_\parallel,t)$ and the total electric field applied $\bm{E}(\bm{r}_\parallel,t)$, through the relation~\cite{GiulianiVignale}:
\begin{equation}\label{eqn:constitutive}
    J_{i}(\bm{r}_\parallel,t) = \int_{0}^{\infty} d\tau\int d\bm{r}_\parallel^\prime \sigma_{ij}(\bm{r}_\parallel,\bm{r}_\parallel^\prime,\tau) E_{j}(\bm{r}_\parallel^\prime,t-\tau)~,
\end{equation}
where summation over repeated indices is implicit.
In a bilayer structure, Eq.~\eqref{eqn:constitutive} must take into account the layer degree of freedom, and in this sense we can recast the constitutive equation for the current and the applied electric field as:
\begin{equation}\label{eqn:constitutive_layer}
    J^{(\ell)}_{i}(\bm{r}_\parallel,t) = \int_{0}^{\infty} d\tau\int d\bm{r}_\parallel^\prime \sigma^{(\ell,\ell^\prime)}_{ij}(\bm{r}_\parallel,\bm{r}_\parallel^\prime,\tau) E^{(\ell^\prime)}_{j}(\bm{r}_\parallel^\prime,t-\tau)~,
\end{equation}
where $\ell,\ell^\prime = 1,2$ are the layer indices. The position vectors are on the $\bm{x}-\bm{y}$ plane, the $\bm{z}$ direction being perpendicular to the layers. In a translationally invariant system, we can Fourier transform with respect to space and time, so that Eq.~\eqref{eqn:constitutive_layer} becomes:
\begin{equation}\label{eqn:constitutive_layer_Fourier}
    J^{(\ell)}_{i}(\bm k,\omega) = \sigma^{(\ell,\ell^\prime)}_{ij}(\bm k,\omega) E^{(\ell^\prime)}_{j}(\bm k,\omega)~.
\end{equation}
More explicitly, the optical conductivity $\sigma^{(\ell,\ell^\prime)}_{ij}(\bm k,\omega)$ is a $4\times4$ tensor and taking a homobilayer can be explicitly written as~\cite{Morell_2DMaterials_2017,stauber_2018,ding_2023}: 
\begin{equation}\label{eqn:constitutive1}
	   \left(\begin{array}{c}
	   J_{x}^{(1)}\\
	   J_{y}^{(1)}\\
	   J_{x}^{(2)}\\
	   J_{y}^{(2)}
	   \end{array}\right)= \left(\begin{array}{cccc}
		\sigma_0  & 0 & \sigma_1 & \sigma_{xy} \\
		0 & \sigma_0 & -\sigma_{xy} & \sigma_1 \\
		\sigma_1 & -\sigma_{xy} & \sigma_0 & 0 \\
		\sigma_{xy} & \sigma_1 & 0 & \sigma_0
	\end{array}\right) \left(\begin{array}{c}
	E_{x}^{(1)}\\
	E_{y}^{(1)}\\
	E_{x}^{(2)}\\
	E_{y}^{(2)}
	\end{array}\right)~.
\end{equation}
We have omitted writing the arguments $\bm{k}, \omega$ everywhere: $J_{x}^{(1)}(\bm{k}, \omega)$, $\sigma_0(\bm{k}, \omega)$, etc. In writing Eq.~\eqref{eqn:constitutive1} we have assumed a $C_n$ ($n\geq3$) rotational invariance in the 2D plane. Every quantity in Eq.~\eqref{eqn:constitutive1} is resolved in the layer index. We stress that the quantities $\sigma_1$ and $\sigma_{xy}$ are in general non-zero also when the inter-layer electron-electron interactions are neglected because of inter-layer tunneling. The quantity $\sigma_{xy}$ comes from the chirality of the twisted bilayer and is a {\it pseudo-scalar} that changes sign upon spatial inversion ($\theta\to-\theta$).
\subsection{Reflection tensor}
The scattering problem of an incident EM field at the chiral bilayer interface,~i.e. determining the reflected and transmitted EM fields, can be solved by usual methods~\cite{Landau08, peres, lin_prl_2020, ho_2023}. The resulting reflection tensor $r^{\alpha,\beta}$ depends on the optical conductivity of the chiral bilayer~\eqref{eqn:constitutive1} and the frequency of the incident field. Here we show results for the reflection tensor $r^{\alpha,\beta}$ expanded to first order in $k_{z,n} d \ll 1$ $n=1,2,3$, meaning that the spacing between the layers is the smallest length scale of the problem (typically of the order of $\approx0.3$ nm~\cite{fukaya_prbl_2021}). We also restrict ourselves to the case of Fig.~\ref{fig1}, fixing the dielectric constants to $\varepsilon_1 = \varepsilon_3 \equiv \varepsilon$ and $\varepsilon_2 = 1$. With this choice of parameters, $k_{1}= k_{3} = \sqrt{\varepsilon}\omega/c$, and $k_{2} = k_0 =\omega/c$. The corresponding transverse wave vectors are obtained from the definition in Eq.~\eqref{eqn:transverse_wavevector}. In particular, we call $k_{z,1} = k_{z,3} \equiv k_{z}$ and $k_{z,2} \equiv k_{0,z}$ for brevity. A detailed calculation is reported in length in the Appendix~\ref{appendixE}. The reflection tensor elements $r^{\alpha,\beta}$ for a chiral bilayer with the conductivity tensor given by Eq.~\eqref{eqn:constitutive} are:
\begin{gather}
    r^{s,s}= r^{s,s}_{\rm SL} -\frac{i d k_{z} }{2}\frac{\Lambda_{\rm T} ^2-c^2 k_{0,z}^2 \left[c^2-16 \pi^2  \sigma_{xy}^2\right]}{\Lambda_{\rm T} ^2}~,\label{eqn:rss}\\
    r^{p,p} = r^{p,p}_{\rm SL} + \frac{i d \varepsilon}{2 k_{z} }\frac{\Sigma_{+}^{(1)} \Sigma_{-}^{(1)}+16 \pi ^2 k_{z}^2 \sigma_{xy}^2 \omega ^2c^{-2}}{\Lambda_{\rm L} ^2}~,\label{eqn:rpp}\\
    r^{p,s} = -\frac{2 i \pi  d \sigma_{xy}\sqrt{\varepsilon }}{c} \frac{ \Sigma^{(2)}}{ \Lambda_{\rm T}\Lambda_{\rm L}}~,\label{eqn:rps}\\
    r^{s,p} = -r^{p,s}~.
\end{gather}
Here we defined for short-hand notation the following quantities:
\begin{gather}
    r^{s,s}_{\rm SL} = \frac{c^2 k_{z}}{\Lambda_{\rm T}}-1~,\\
    r^{p,p}_{\rm SL} = 1-\frac{\omega  \varepsilon}{\Lambda_{\rm L}}~,
\end{gather}
and:
\begin{gather}
    \Lambda_{\rm T} = 4\pi \omega(\sigma_0 + \sigma_1) + c^2 k_{z}~,\\
    \Lambda_{\rm L} = 4\pi k_{z}(\sigma_0 + \sigma_1) + \omega\varepsilon~,\\
    \Sigma_{\pm}^{(1)} = 4\pi k_{z}k_{0,z}(\sigma_0 + \sigma_1) \pm k_{z}\omega + k_{0,z}\omega\varepsilon~,\\
    \Sigma^{(2)} = 4 \pi  k_{z}c^2 k^2 (\sigma_{0}+\sigma_{1})  - c^2k_{0,z}^2 \omega  \varepsilon   + c^2k_{z}^2 \omega~.
\end{gather}
In writing Eqs.~\eqref{eqn:rss}-\eqref{eqn:rps} we retained the contributions proportional to $\sigma_{xy}$ and its powers explicit. 

We can already infer two important facts: i) the diagonal coefficients, i.e. the $r^{s,s}$ and $r^{p,p}$ terms, are modified by the chiral contribution to the optical conductivity $\sigma_{xy}$, but they do not depend on its sign as it appears in the numerator every time squared; ii) the off diagonal elements, i.e. $r^{s,p}$ and $r^{p,s}$, are nonzero only if $\sigma_{xy}\neq0$ and they are sensible to its sign (i.e. to the chirality of the substrate). In view of calculating the total dissipated power, we stress that, as we already noticed, thanks to the chiral symmetry of the twisted bilayer interface, $r^{s,p} = -r^{p,s}$ so that $P_{e,m}^{\rm ref} = P_{m,e}^{\rm ref}$ (Eqs.~\eqref{eqn:Pem} and~\eqref{eqn:Pme}), and they add up when summing all the contributions in Eq.~\eqref{eqn:total_dissipated_power}.

Finally, we take here the time to notice that, since the change in handness of the twisted bilayer ($\theta\to-\theta$) changes the sign of both $r^{p,s}$ and $r^{s,p}$, from Eq.~\eqref{eqn:Pem} (the same can be said for $P_{m,e}^{\rm ref}$, Eq.~\eqref{eqn:Pme}) we immediately infer for $P_{e,m}^{\rm ref}$ that:
\begin{equation}
    P_{e,m}^{\rm ref}(\bm{d}_\pm,\pm\theta) = -P_{e,m}^{\rm ref}(\bm{d}_\mp,\pm\theta) = -P_{e,m}^{\rm ref}(\bm{d}_\pm,\mp\theta)~.
\end{equation}
\subsection{Chiral dissymetry factor}
We quantify the strength of the effect of the substrate on the fluorescence decay of the chiral molecule, by introducing the decay rate dissymetry factor $g$:
\begin{equation}\label{eqn:g_def}
    g = \frac{|\Gamma_{+} - \Gamma_{-}|}{\Gamma_{+} + \Gamma_{-}}~,
\end{equation}
where $\Gamma_{+}$ and $\Gamma_{-}$ are the decay rates associated to $\bm{d}_{+}$ and $\bm{d}_{-}$ for a fixed substrate chirality, respectively, and are obtained combining Eq.~\eqref{eqn:Gamma_P} and Eq.~\eqref{eqn:total_P}.
The factor $g$ quantifies the relative difference in decay rate for the two handnesses of the chiral molecules.
\section{Numerical Results}\label{sectionIV}
\begin{figure*}
    \centering
    \begin{tabular}{ll}
    \begin{overpic}[width=0.5\textwidth]{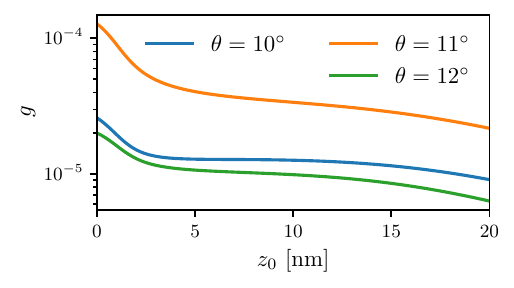}%
    \put(0,54){(a)}%(665,550)
    \end{overpic}
    \begin{overpic}[width=0.5\textwidth]{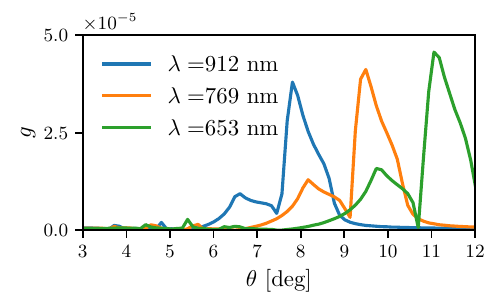}%
    \put(0,54){(b)}%(665,550)
    \end{overpic}
    \end{tabular}
    \caption{(a) Dependence of the decay rate dissymmetry factor $g$ on the distance $z_0$ between the HBNG1 molecule and the substrate. These results have been obtained by fixing the frequency to $\omega=1.896~{\rm eV}$, corresponding to a wavelength of $\lambda\approx 653~{\rm nm}$. (b) Dependence of the decay rate dissymmetry factor $g$ on the twist angle $\theta$ of TBG. The distance between HBNG1 and TBG is fixed to $z_0=4~{\rm nm}$.}
    \label{fig2}
\end{figure*}

In this Section we present the main numerical results. We consider TBG as the chiral twisted bilayer interface. Details on the continuum model and the Kubo formula for the layer resolved conductivity tensor appearing in Eq.~\eqref{eqn:constitutive1} are reported in Appendix~\ref{appendixF}. 
As a case of study, we have chosen helical bilayer nanographenes (HBNG1)~\cite{yang_2023} as the quantum emitter. We conducted accurate first-principles calculations in order to obtain the electric and magnetic moments associated to the principal decay channel of HBNG1.

Finally, we fix the dielectric constants of regions 1 and 3 to $\varepsilon_1 = \varepsilon_3 \equiv \varepsilon = 4.9$, simulating an hexagonal boron nitrate (hBN) environment. All calculations are performed at room temperature $T=300$ K and for {\it undoped} TBG.
\subsection{DFT calculation}
Optimizations of the HBNG1 molecules in the ground state were conducted utilizing Density Functional Theory (DFT) as implemented in the ORCA ab initio quantum chemistry package~\cite{Neese_2020}. We employed the Becke-3-parameter-Lee–Yang–Parr (B3LYP) exchange-correlation functional with the basis set 6-311G(2d,p). All the geometry optimisations were done in the gas phase and based on the single crystal structure. Given that in typical experiments~\cite{moreno_2024}, the chiral layered substrate (TBG) is embedded in an heterostructure that contains hBN in the first layer near to the molecule, we have performed further optimizations of the HBNG1, previously relaxed with ORCA, on hBN substrates. With this aim, the Vienna Ab Initio Simulation Package (VASP) was used employing the Perdew-Burke-Ernzerhof (PBE) generalized gradient approximation (GGA) to deal with the electron exchange-correlation interaction and projected augmented wave pseudo-potentials to describe the ion-electron interaction. We considered a unit cell formed by a single HBNG1 molecule in proximity (by a distance of $\approx 2.5 {\rm \AA}$ measured along the chiral axis of the molecule) to a slab of $14\times 14\times 1$ of hBN. We used a $1\times 1\times 1$ k - points mesh for the calculations.

Following the geometry optimizations, we calculated the electronic transition chiro-optical properties of the (further relaxed) HBNG1 molecule with ORCA by Time Dependent Density Functional Theory (TD-DFT) using a B3LYP exchange-correlation functional with the gaussian basis set 6-311G(2d,p). TD-DFT calculations suggest that the lowest-energy transitions occur at wavelengths of $653.9$nm and $582.7$nm for oscillator strengths of  $f=0.029$ and $f=0.092$, respectively, and correspond both to the transitions H$\rightarrow$L and H$\rightarrow$L+1, which can be compared to the experimental and calculated UV-Vis spectra reported in reference~\cite{yang_2023}. Table\,\ref{TDM} summarizes the results for the transition electric and magnetic dipoles for the first two excited-to-ground state transitions.
\begin{widetext}
\begin{center}
\begin{table} 
%\centering
\begin{tabular}{c c c c c c}
\hline \hline
 State & Wavelength (nm) & \quad  $\mu_0$ ($10^{-20}$ esu-cm) & \quad  $m_0$ ($10^{-20}$ erg/Gauss) & \quad $\cos(\varphi)$ & \quad $\mathcal{G}$ \\
\hline
  1 & 653.9 & \quad  200.961 & \quad 1.261 & \quad -0.099 & \quad 0.0025 \\
  2 & 582.7 & \quad 337.017 &  \quad 2.566 & \quad 0.673 & \quad -0.0204 \\
\hline \hline
\end{tabular}
\caption{Calculated transition dipole moments, angle between them, and intrinsic dissymmetry factor, for the first two excited-to-ground state transitions of the HBNG1 molecule.}
\label{TDM}
\end{table}
\end{center}
\end{widetext}

\subsection{SE results}
We present here the main results concerning the SE decay rate. All the calculations have been performed taking the near field approximation,~i.e. $\omega/c\ll k$ so that $k_{z} = k_{z,0} \approx ik$. This approximation is justified if the distance between the emitter and the substrate is much smaller than the wavelength of the emitted light $z_0\ll \lambda$, with $\lambda = 2\pi c/\omega$. In this limit the contributions induced by the TBG substrate to the total dissipated power take the following form:
\begin{gather}
    \frac{P_{e,e}^{\rm ref}(z_0,\lambda,\theta)}{P_0} = \frac{3}{2}\frac{\lambda^3}{z_0^3}{\cal F}_{e,e}(z_0,\lambda,\theta)~, \label{eqn:Pee_NF}\\
    \frac{P_{e,m}^{\rm ref}(z_0,\lambda,\theta)}{P_0} = \pm\frac{3}{4}\frac{m_0}{\mu_0}\frac{\lambda^4}{z_0^4}{\cal F}_{e,m}(z_0,\lambda,\theta)~, \label{eqn:Pem_NF}\\
    \frac{P_{m,m}^{\rm ref}(z_0,\omega,\theta)}{P_0} \approx 0~,\label{eqn:Pmm_NF}
\end{gather}
where the dimensionless form factors ${\cal F}_{e,e}(z_0,\lambda,\theta)$ and ${\cal F}_{e,m}(z_0,\lambda,\theta)$ are defined by:
\begin{equation}
    {\cal F}_{e,e}(z_0,\lambda,\theta) \equiv \int_0^\infty \frac{ds s^3}{(2\pi)^3} e^{-2s}\Im \left[\frac{2 i(\sigma_0 + \sigma_1)}{2 i s(\sigma_0 + \sigma_1) + \frac{z_0}{\lambda}c\varepsilon}\right]~,\\
\end{equation}
\begin{equation}
\begin{split}
    &{\cal F}_{e,m}(z_0, \lambda,\theta) \equiv \frac{d}{\lambda}\int_0^\infty \frac{ds s^3}{(2\pi)^3} e^{-2s}\times\\
    &\Im \left[\frac{4\pi i\varepsilon\sigma_{xy}}{c}\frac{i(\varepsilon-1)\frac{z_0}{\lambda} - 2 s(\sigma_0 + \sigma_1)c^{-1}}{i\varepsilon\frac{z_0}{\lambda} - 2 s(\sigma_0 + \sigma_1)c^{-1}}e^{i\varphi}\right]~,
\end{split}
\end{equation}
and thanks to the symmetries of the substrate $P_{m,e}^{\rm ref}(z_0,\lambda,\theta) = P_{e,m}^{\rm ref}(z_0,\lambda,\theta)$.
Using these expressions, namely Eqs.~\eqref{eqn:Pee_NF},~\eqref{eqn:Pem_NF} and~\eqref{eqn:Pmm_NF}, we obtain the chiral dissymetry factor $g$, defined in Eq.~\eqref{eqn:g_def}:
\begin{equation}\label{eqn:g_NF}
    g(z_0,\lambda, \theta) = \frac{m_0}{\mu_0}\frac{|{\cal F}_{e,m}(z_0,\lambda,\theta)|}{\frac{2}{3}\frac{z_0^4}{\lambda^4} + \frac{z_0}{\lambda}{\cal F}_{e,e}(z_0,\lambda,\theta)}~.
\end{equation}
%
%Eq.~\eqref{eqn:g_NF} shows that, in the geometry considered, $g(z_0,\lambda, \theta)$ is directly related to the ratio between the modulus of the magnetic and electric dipole $m_0/\mu_0$. 
Figure~\ref{fig2}(a) illustrates the variation of $g(z_0, \lambda, \theta)$ as a function of $z_0$ for $\lambda = 653~{\rm nm}$, the wavelength corresponding to the primary optical transition of HBGN1, and for a few values of the twist angle $\theta$ in TBG. Two distinct behaviors can be observed: at short distances, both ${\cal F}_{e,m}(z_0, \lambda, \theta)$ and $z_0{\cal F}_{e,e}(z_0, \lambda, \theta)$ exhibit a strong dependence on $z_0$, resulting in the dissymmetry factor reaching its maximum as $z_0 \to 0$. At longer distances, ${\cal F}_{e,m}(z_0, \lambda, \theta)$ and $z_0{\cal F}_{e,e}(z_0, \lambda, \theta)$ approach a constant value, leading to $g(z_0, \lambda, \theta)$ scaling as $z_0^{-4}$~\cite{gaudreau_nano_lett_2013}. This behavior remains consistent even as the twist angle $\theta$ is varied. Figure~\ref{fig2}(b) depicts the dependence of the dissymmetry factor on the twist angle for different wavelengths of emitted light, with a fixed molecule-substrate distance of $z_0 = 4~{\rm nm}$. In particular, Figure~\ref{fig2}(b) reveals that $g(z_0, \lambda, \theta)$ exhibits a pronounced peak around an optimal twist angle for a given wavelength $\lambda$. This implies that for a specific emission wavelength characteristic of the molecule, an optimal substrate can always be engineered by adjusting the twist angle of the chiral bilayer.

\section{Summary and conclusions}
\label{sect:summary}
In this work, we explored the potential of using chiral bilayer substrates to modify the spontaneous emission (SE) of chiral molecules. Through theoretical analysis, we demonstrated that a chiral bilayer substrate can induce significant and non-trivial contributions to the total decay rate of a chiral quantum emitter. We focused on twisted bilayer graphene (TBG) as the chiral bilayer and HBNG1 as the chiral molecule, conducting detailed numerical analyses.

Specifically, we employed optimized ab initio calculations to determine the electronic transition chiro-optical properties of HBNG1. Additionally, we treated TBG within the continuum model and show that at short molecule-substrate distances, the decay rate dissymmetry factor is highly sensitive to the molecule’s proximity, reaching its maximum as the distance approaches zero $z_0\to0$. At longer distances, the SE modification exhibits characteristic scaling behavior, with the dissymmetry factor decreasing as $z_0^{-4}$, a trend that remains consistent across various twist angles.

Moreover, we identified an optimal twist angle for each characteristic emission wavelength, where the dissymmetry factor peaks. These findings open up possibilities for tailoring substrate optical properties to enable fluorescence-based chiral sensing. Our work suggests that chiral bilayers offer a promising platform for enantiomeric discrimination, with the potential to significantly enhance the sensitivity of chirality detection in molecular systems.

In conclusion, we proposed a novel approach to exploiting the chirality of quantum emitters by asymmetrically modifying their fluorescence properties, laying the groundwork for future advances in nanoscale chiral sensing.
\begin{acknowledgments}
This work was supported by i) the European Union's Horizon 2020 research and innovation programme under grant agreement no.~881603 - GrapheneCore3 and the Marie Sklodowska-Curie grant agreements No.~873028 (HYDROTRONICS) and No.~101131579 (EXQIRAL) and ii) the Italian Minister of University and Research (MUR) under the ``Research projects of relevant national interest  - PRIN 2020''  - Project no.~2020JLZ52N, title ``Light-matter interactions and the collective behavior of quantum 2D materials (q-LIMA)''. P.J.H. acknowledges support by the Gordon and Betty Moore Foundation’s EPiQS Initiative through grant GBMF9463, the Fundacion Ramon Areces, and the ICFO Distinguished Visiting Professor program. F.H.L.K. acknowledges financial support from  the ERC TOPONANOP  (726001), the Government of Catalonia trough the SGR grant, the Spanish Ministry of Economy and Competitiveness through the Severo Ochoa Programme for Centres of Excellence in R\&D (Ref. SEV-2015-0522) and Explora Ciencia (Ref. FIS2017- 91599-EXP), Fundacio Cellex Barcelona, Generalitat de Catalunya through the CERCA program, the Mineco grant Plan Nacional (Ref. FIS2016-81044-P), and the Agency for Management of University and Research Grants (AGAUR) (Ref. 2017-SGR-1656).
\end{acknowledgments}
%

%Appendix

\setcounter{section}{0}
\setcounter{equation}{0}%
\setcounter{figure}{0}%
\setcounter{table}{0}%

\renewcommand{\thetable}{A\arabic{table}}
\renewcommand{\theequation}{A\arabic{equation}}
\renewcommand{\thefigure}{\thesection.\arabic{figure}}% Figure counter representation
\renewcommand{\theHfigure}{\thesection.\arabic{figure}}% Hyperref figure hyperlink hook
\renewcommand{\bibnumfmt}[1]{[#1]}
\renewcommand{\citenumfont}[1]{#1}

\onecolumngrid
\newpage
\numberwithin{equation}{section}
%
%\begin{appendices}
%
\appendix
\setcounter{page}{1}
\section{The dipole approximation and the dissipated power picture}\label{appendixA}
\subsection{The dipole approximation}
When treating a quantum system that interacts with light, we can often simplify the analysis by using a two-level model. In this model, the emission properties of the system can be understood through the macroscopic theory of the power dissipated by the electromagnetic (EM) dipole associated with the transition between the two levels. This method is commonly known as the ``{\it dipole approximation}''~\cite{app_novotny}.

The emission properties of a chiral molecule are a result of the interplay between electric and magnetic transitions~\cite{app_meskers_2022}. In fact, for the allowed transitions of chiral molecules, the degree of circular polarization in luminescence emission can be substantial, indicating a departure from the well-known and commonly used electric dipole approximation~\cite{app_meskers_2022}. The combination of electric and magnetic dipoles captures both the linear and circular aspects of the charge's motion during the spectroscopic transition in helical structures. We then introduce the oscillating {\it chiral} dipole describing the principal optical transition with energy $\hbar\omega$ as~\cite{app_eismann_optica_2018} (we assume a harmonic time dependence):
\begin{equation}
    \bm{d}_{\pm}(\bm{r}, t) = (\pm i \bm{\mu} + \bm{m})\delta(\bm{r} -\bm{r}_0)e^{-i\omega t}~,
\end{equation}
i.e. a dipole that consists of an electric $\bm{\mu}\in\mathbb{R}^3$ and a magnetic $\bm{m}\in\mathbb{R}^3$ dipole moment with relative phase of $\pm \pi/2$. The sign $\pm$ identifies the chirality of the molecule and defines the helicity of the light emitted in the far field~\cite{app_fernandez-corbaton_2013}. Furthermore, the dipole is supposed to be located at $\bm{r} =\bm{r}_0$. 
\subsection{Decay rate}
The dipole approximation described above can be used to model how the fluorescence decay rate of a molecule is modified by an inhomogeneous environment. In particular, when the free-space average $P_0$ and the total $P$ dissipated powers are known, one can obtain the decay rate $\Gamma$ of the transition in the hypothesis that all the energy dissipated by the oscillator is transformed into radiation~\cite{app_novotny}:
\begin{equation}\label{eqn:decay_rate_relation_P}
    \frac{\Gamma}{\Gamma_0} = \frac{P}{P_0}~,
\end{equation}
where $\Gamma_0$ is the free-space decay rate. Since we are dealing with electric and magnetic dipoles, $\Gamma_0$ is the superposition of the two free-space contributions:
\begin{equation}\label{eqn:gamma0}
    \Gamma_0 = \Gamma_{0,e} + \Gamma_{0,m} = \frac{4\omega^3}{3\hbar c^3}(|\bm{\mu}|^2 + |\bm{m}|^2)~.
\end{equation}
Eq.~\eqref{eqn:decay_rate_relation_P} relates the quantum-mechanical picture of spontaneous emission (the left-hand side) to the classical formalism of dipole radiation (the right-hand side). Due to the relationship of
Eq.~\eqref{eqn:decay_rate_relation_P} we can classically calculate the relative change of the spontaneous decay rate of a quantum emitter in an inhomogeneous environment~\cite{app_novotny,app_koppens_nano_letters_2011, app_kort-kamp_prb_2015}. From now on, we will focus on the calculation of the dissipated power $P$, intending that, owing to Eq.~\eqref{eqn:decay_rate_relation_P}, once $P$ is known, the total decay rate $\Gamma$ is also known. 

Since the chiral molecule is modeled as a superposition of electric and magnetic dipoles, the total dissipated power is a combination of two contributions: the electric $P_{e}$ and the magnetic $P_{m}$ contribution, describing how the electric and magnetic dipoles lose energy, respectively. Each of these terms contains a ``diagonal'' and a ``cross'' term. In total we have four distinct quantities, namely the electric-electric, magnetic-magnetic, magneto-electric, and electro-magnetic dissipated powers:
\begin{equation}
    P = P_{e} + P_{m} = P_{e,e} + P_{m,m} + P_{e,m} + P_{m,e}~.
\end{equation}
The four contributions to the total dissipated power are given by~\cite{app_novotny}:
\begin{gather}
    P_{e,e}(\bm{r}_0,\omega) = \frac{\omega}{2}\Im \left[\left(\pm i \mu_i\right)^* E_{\mu,i}(\bm{r}_0,\omega)\right]~,\label{eqn:dissipated_power1} \\
    P_{e,m}(\bm{r}_0,\omega) = \frac{\omega}{2}\Im \left[\left(\pm i \mu_i\right)^* E_{m,i}(\bm{r}_0,\omega)\right]~,\label{eqn1:dissipated_power2} \\
    P_{m,e}(\bm{r}_0,\omega) = \frac{\omega}{2}\Im \left[m_i^* H_{\mu,i}(\bm{r}_0,\omega)\right]\label{eqn:dissipated_power3} ~,\\
    P_{m,m}(\bm{r}_0,\omega) = \frac{\omega}{2}\Im \left[m_i^*  H_{m,i}(\bm{r}_0,\omega)\right]\label{eqn:dissipated_power4}~,
\end{gather}
In Eqs.~\eqref{eqn:dissipated_power1}-\eqref{eqn:dissipated_power4} the symbol $*$ denotes the complex conjugation. 
The EM fields $\bm{E}_\mu, \bm{H}_\mu$ and $\bm{E}_m, \bm{H}_m$ are generated by the electric and magnetic dipoles, respectively, and are evaluated at the dipole position $z =z_0$. Notice that these fields are the total fields in the region $z>0$ and contain both the {\it free} and {\it reflected} contributions. In other terms, the electric and magnetic dipoles produce both an electric and a magnetic field. These fields can be altered by the various conditions at the boundary of the environment where the chiral dipole $\bm{d}_{\pm}$ is placed. The EM field reflected by the interface at $z=0$ (see Fig.~\ref{fig1}) that ``returns'' to the dipole causes a modification to the free-space power dissipation. So, in order to describe the total energy dissipation of the dipole, the electric field to be assessed is a combination of the fields emitted by the dipole itself and the scattered fields from the environment. 
\subsection{Dyadic Green function approach}
In order to evaluate the total dissipated power, we describe the total EM field in the region of space where the dipole $\bm{d}_\pm$ is placed. A simple, yet powerful, method for calculating the EM field in an inhomogeneous environment is provided by the Dyadic Green functions (DGfs) approach to electromagnetism~\cite{app_novotny, app_Tai1994, app_peres}. The electric $G_{ik}(\bm{r},\bm{r}^\prime,\omega)$ and magnetic $G_{ik}^{\rm M}(\bm{r},\bm{r}^\prime,\omega)$ DGfs are the physical response functions that allow one to express the electric $E_i({\bm r}, \omega)$ and magnetic $H_i({\bm r}, \omega)$ fields at any point ${\bm r}$ in space induced by an external current ${\bm J}({\bm r}^\prime,\omega)$ located at a point ${\bm r}^\prime$. (In Appendix~\ref{appendixB} the reade can find further details on this formalism). Once the DGfs are known, the EM field generated by the chiral dipole $\bm{d}_\pm$ can be obtained in all space:
\begin{gather}
    E_{\mu,i}(\bm{r},\omega) = \frac{\omega^2}{c^2} G_{ij}(\bm{r},\bm{r}_0,\omega) (\pm i \mu_j)~,\\
    H_{\mu,i}(\bm{r},\omega) = \frac{-i \omega}{c}\epsilon_{ijk}\partial_j G_{k\ell}(\bm{r},\bm{r}_0,\omega) (\pm i \mu_\ell)~,\\
    E_{m,i}(\bm{r},\omega) = \frac{i\omega}{c}\epsilon_{ijk}\partial_{j} G_{k\ell}^{\rm M}(\bm{r},\bm{r}_0,\omega) m_{\ell}~,\\
    H_{m,i}(\bm{r},\omega) = \frac{\omega_0^2}{c^2} G_{ij}^{\rm M}(\bm{r},\bm{r}_0,\omega) m_{j}~.
\end{gather}
We notice that the DGfs can be divided in three contributions~\cite{app_novotny}: the free-space part $G_{0,ij}$, associated with the primary dipole field; the reflected part $G_{{\rm ref},ij}^{({\rm M})}$, associated with the field scattered by the interface in the upper semi-space; and the transmitted part $G_{{\rm tr},ij}^{({\rm M})}$ associated to the field that passes through the interface in the lower semi-space. Furthermore, while $G_{0,ij}$ gives the free-space contribution to the dissipated power $P_0$, the reflection DGf $G_{{\rm ref},ij}^{({\rm M})}$ embodies the effects of the environment on the semi-space where the dipole is located and is related to the reflection coefficients of the interface itself~\cite{app_novotny} (see also Appendix~\ref{appendixC} for more details). 
Thanks to this subdivision, we can further handle the dissipated powers in Eqs.~\eqref{eqn:dissipated_power1}-\eqref{eqn:dissipated_power4}. As already mentioned, the free-space DGf gives the dissipated power in free-space $P_0 = \frac{\hbar\omega}{4} \Gamma_0$~\cite{app_novotny}, where $\Gamma_0$ is given by~\eqref{eqn:gamma0}, while the reflected part gives the correction due to the environment:
\begin{equation}\label{eqn:total_dissipated_power}
    P = P_0 + P_{e,e}^{\rm ref} + P_{m,m}^{\rm ref} + P_{e,m}^{\rm ref} + P_{m,e}^{\rm ref}~,
\end{equation}
where
\begin{gather}
    P_{e,e}^{\rm ref}(z_0,\omega) = \frac{\omega^3}{2c^2}\int\frac{d^2\bm{k}}{(2\pi)^2}\Im \left[\mu_i  G_{{\rm ref},ij}(\bm{k},z_0,z_0,\omega) \mu_j\right]~,\label{eqn:ref_dissipated_power1} \\
    P_{m,m}^{\rm ref}(z_0,\omega) = \frac{\omega^3}{2c^3}\int\frac{d^2\bm{k}}{(2\pi)^2}\Im \left[m_i G_{{\rm ref},ij}^{\rm M}(\bm{k},z_0,z_0,\omega) m_{j}\right]\label{eqn:ref_dissipated_power4}~,\\
    P_{e,m}^{\rm ref}(z_0,\omega) = \pm \frac{\omega^2}{2c}\int\frac{d^2\bm{k}}{(2\pi)^2}\Im \left[\mu_i \epsilon_{ijk}\left.\partial_{j} G_{{\rm ref},k\ell}^{\rm M}(\bm{k},z_0,z_0,\omega)\right|_{z=z_0} m_{\ell}\right]~,\label{eqn:ref_dissipated_power2} \\
    P_{m,e}^{\rm ref}(z_0,\omega) = \pm\frac{\omega^2}{2c}\int\frac{d^2\bm{k}}{(2\pi)^2}\Im \left[m_i\epsilon_{ijk}\left.\partial_j G_{{\rm ref},k\ell}(\bm{k},z_0,z_0,\omega) \right|_{z=z_0}\mu_\ell\right]\label{eqn:ref_dissipated_power3}~,
\end{gather}
are the explicit expressions of the dissipated power due to the scattering at the interface, and since the system has cylindrical symmetry due to the translational invariance along the interface plane, we introduced the Fourier transform of the DGfs:
\begin{equation}
    G_{ij}^{({\rm M})}({\bm r}_\parallel,{\bm r}_\parallel^\prime,z,z^\prime,\omega) =\int\frac{d\bm k}{(2\pi)^2} e^{i\bm{k}\cdot(\bm{r}_\parallel-\bm{r}_\parallel^\prime)} G_{ij}^{({\rm M})}(\bm k,z,z^\prime,\omega)~.
\end{equation}
The full expressions of the reflected electric $G_{{\rm ref},ij}(\bm{k},z,z^\prime,\omega)$ and magnetic $G_{{\rm ref},ij}^{\rm M}(\bm{k},z,z^\prime,\omega)$ DGf are derived in Appendix~\ref{appendixC}.
\section{Dyadic Green's functions}\label{appendixB}
The calculations presented in the main text rely on the formulation of electrodynamics through the dyadic Green's function formalism. Maxwell equations in linear materials, with local electric and magnetic permeabilities are, in CGS units:
\begin{equation}
\begin{cases}
\nabla \cdot{\bm D}(\bm r, \omega) = 4\pi \rho_{\rm ext} (\bm r, \omega)~,\\
\nabla \cdot{\bm  B }(\bm r, \omega) = 0~,\\
\nabla \times{\bm E}(\bm r, \omega) = \frac{i\omega}{c} {\bm B} (\bm r, \omega)~,\\
\nabla \times {\bm H} (\bm r, \omega) = \frac{4\pi}{c} {\bm J}_{\rm ext}(\bm r, \omega)-\frac{i\omega}{c} {\bm D} (\bm r, \omega)~.
\end{cases}
\end{equation}
The electric displacement $\bm{D}(\bm{r},\omega)$ and magnetic induction $\bm{B}(\bm{r},\omega)$ fields are related to the electric and magnetic fields by the linear and local relationships 
\begin{align}
{\bm D}(\bm r, \omega) & =   \varepsilon(\bm r, \omega)\cdot {\bm E} (\bm r, \omega),\label{eq:epsilon}\\
{\bm  B }(\bm r, \omega)& = \mu(\bm r, \omega)\cdot  {\bm H} (\bm r, \omega) ,\label{eq:mu}
\end{align}
with $ \epsilon(\bm r, \omega)$ and $\mu(\bm r, \omega)$ are, in general, position and frequency dependence tensors.

Maxwell equations imply that the electric and magnetic fields satisfy the inhomogeneous Helmoltz equation:
\begin{equation}
\nabla \times \{\mu^{-1} (\bm r, \omega)\cdot[ \nabla \times{\bm E}(\bm r, \omega)]\}-\frac{\omega^2}{c^2} \varepsilon(\bm r, \omega)\cdot {\bm E} (\bm r, \omega) = \frac{4\pi i \omega}{c^2} {\bm J}_{\rm ext}(\bm r, \omega). \label{eq:HelE}
\end{equation}
\begin{equation}
\nabla \times\{  \varepsilon^{-1}(\bm r, \omega)\cdot [\nabla \times {\bm H} (\bm r, \omega)]\}-\frac{\omega^2}{c^2}\mu(\bm r, \omega)\cdot  {\bm H} (\bm r, \omega) = \frac{4\pi}{c} \nabla \times  [\varepsilon^{-1}(\bm r, \omega)\cdot {\bm J}_{\rm ext}(\bm r, \omega)].\label{eq:HelH}
\end{equation}

We introduce the following compact notation for a differential operator:
\begin{align}\label{eq:magnetic_differential_operator}
{\cal D}_{ij}[{\cal M}]&\equiv \epsilon_{ink}\epsilon_{\ell m j}\partial_n\left[{\cal M}_{k\ell}({\bm r}, \omega)\partial_m\right]~.
\end{align}
In Eq.~(\ref{eq:magnetic_differential_operator}), ${\cal M} = {\cal M}_{ij}({\bm r},\omega)$ is a generic space- and frequency-dependent rank-$2$ tensor.
\begin{equation}
{\cal D}_{ij}(\alpha) \equiv \{\nabla \times [\alpha(\bm r, \omega)\cdot\nabla \times]\}_{ij} =\epsilon_{ik\ell}\epsilon_{mnj}\partial_k[\alpha_{\ell m}(\bm r, \omega)\partial_n \dots],\label{eq:doperator}
\end{equation}
in the simple case $\alpha_{ij}(\bm r, \omega) = \delta_{ij}$ this reduces to the well known identity for the curl of the curl ($\bm{\nabla}\times\bm{\nabla}\times$)
\begin{equation}
{\cal D}_{ij}(\bm 1) =\partial_i\partial_j-\delta_{ij}\partial_k\partial_k~.\label{eq:rotrot}
\end{equation}
With this notation the Helmoltz equations can be written as
\begin{align}
{\cal D}_{ij}(\mu^{-1})E_j(\bm r, \omega)-\frac{\omega^2}{c^2} \varepsilon_{ij}(\bm r, \omega) E_j (\bm r, \omega)& = \frac{4\pi i \omega}{c^2}  J_{{\rm ext},i}(\bm r, \omega)~,\\
{\cal D}_{ij}(\varepsilon^{-1})H_j (\bm r, \omega)-\frac{\omega^2}{c^2}\mu_{ij}(\bm r, \omega) H_j (\bm r, \omega) & = \frac{4\pi}{c} \epsilon_{ijk} \partial_j[\varepsilon^{-1}_{k\ell}(\bm r, \omega)J_{{\rm ext},\ell}(\bm r, \omega)]~.
\end{align}
Given a differential equation, it is customary to introduce the Green's function associated with the associated linear differential operator, in our case ${\cal L} = {\cal D}_{ij}(\alpha) - \frac{\omega^2}{c^2} \bar{\alpha}_{ij}$. The electric $G_{ik}(\bm{r},\bm{r}^\prime,\omega)$ and magnetic $G_{ik}^{\rm M}(\bm{r},\bm{r}^\prime,\omega)$ Dyadic Green's functions (DGfs) obey the following partial differential equations~\cite{app_Tai1994}:
\begin{equation}\label{eqn:dyadic_electric}
    \left[{\cal D}_{jk}^\prime({}^t\mu^{-1}) - \frac{\omega^2}{c^2} {}^t\varepsilon_{jk}\right]G_{ik}(\bm{r},\bm{r}^\prime,\omega) = 4\pi \delta_{ij}\delta(\bm{r}-\bm{r}^\prime)~,
\end{equation}
\begin{equation}\label{eqn:dyadic_magnetic}
    \left[{\cal D}_{jk}^\prime({}^t\varepsilon^{-1}) - \frac{\omega^2}{c^2} {}^t\mu_{jk}\right]G_{ik}^{\rm M}(\bm{r},\bm{r}^\prime,\omega) = 4\pi \epsilon_{ji\ell}\partial_i\left[\delta_{\ell k}\delta(\bm{r}-\bm{r}^\prime)\right]~.
\end{equation}
In Eqs.~\eqref{eqn:dyadic_electric} and~\eqref{eqn:dyadic_magnetic}, the prime symbol in ${\cal D}^\prime_{jk}$ denotes that the spatial derivatives in Eq.~(\ref{eq:magnetic_differential_operator}) must be carried out with respect to ${\bm r}^\prime$ and ${}^t\mu^{-1}$ (${}^t\varepsilon$)  denotes the transpose of $\mu^{-1}$ ($\varepsilon$). 

A particular solution for the electric field will be given by the convolution of $G_{ij}(\bm{r},\bm{r}^\prime,\omega)$ with the external current source:
\begin{equation}\label{eqn:electric_green_source}
\begin{split}
E_i(\bm r,\omega) & = \frac{i\omega}{c^2}\int_\Omega d\bm r' G_{ij}(\bm r,\bm r',\omega) J_{{\rm ext}, j}(\bm r', \omega)~.
\end{split}
\end{equation}
This equation is known as \textit{volume integral equation}~\cite{app_novotny}: $\Omega$ is the space where the source and the electric field live. We are neglecting boundary terms, thinking at the system with the boundary posed at infinity.
Since $\bm{\nabla}\times\bm{E}(\bm{r},\omega) = \frac{i\omega}{c}\bm{B}(\bm{r},\omega)$, we immediately see that an expression similar to~\eqref{eqn:electric_green_source} can be cast for the magnetic field:
\begin{equation}
    B_i(\bm{r},\omega) = \frac{1}{c}\epsilon_{ijk}\int_\Omega d\bm r' \partial_j G_{jk}(\bm r,\bm r',\omega) J_{{\rm ext}, k}(\bm r', \omega)~.
\end{equation}
In the sections that follow, we will consider a non-magnetic environment, i.e. $\mu_{ij}(\bm{r},\omega)=\delta_{ij}$. Moreover, the dielectrics will be treated as uniform and isotropic $\varepsilon_{ij}(\bm{r},\omega) = \delta_{ij}\varepsilon$. In this particular situation, that is when the space is uniform and isotropic with dielectric constant $\varepsilon$, the solution of the Green's function equation is (once the Fourier transform with respect to $\bm{r}$ is taken)~\cite{appendix_LandauIX}:
\begin{equation}
    G_{ij}(\bm k, \omega) = \frac{4\pi}{k^2 - \frac{\omega^2}{c^2}\varepsilon}\left(\delta_{ij} - \frac{c^2}{\omega^2}\frac{k_i k_j}{\varepsilon}\right)~.
\end{equation}
\section{Derivation of the reflection Dyadic Green's functions}\label{appendixC}
In this Appendix, we derive the expression for the reflection DGfs. This contribution comes from the decomposition of the DGfs in free, reflected and transmitted parts:
\begin{equation}
G_{ij}^{({\rm M})}(\bm r,\bm r',\omega) =
    \begin{cases}
        G_{0,ij}^{({\rm M})}(\bm r,\bm r',\omega) + G_{{\rm ref},ij}^{({\rm M})}(\bm r,\bm r',\omega)~, \quad z>0 \\
        G_{{\rm tr},ij}^{({\rm M})}(\bm r,\bm r',\omega)~, \quad z<0
    \end{cases}
\end{equation}
where the interface between the two semi-spaces is posed at $z=0$. As an intermediate step, we introduce the Fourier transform of the Green's function with respect to the in-plane space coordinates $\bm{r}_\parallel$, $\bm{r}_\parallel^\prime$: 
\begin{equation}
    G_{ij}^{({\rm M})}({\bm r}_\parallel,{\bm r}_\parallel^\prime,z,z^\prime,\omega) = \int\frac{d\bm k}{(2\pi)^2} e^{i\bm{k}\cdot(\bm{r}_\parallel-\bm{r}_\parallel^\prime)} G_{ij}^{({\rm M})}(\bm k,z,z^\prime,\omega)~.
\end{equation}
First we derive the well known expression for the electric reflected DGf~\cite{app_peres,app_novotny}:
\begin{equation}\label{eqn:app_ref_G}
    \bar{\bar{G}}_{\rm ref}(\bm{k},z,z^\prime) = \frac{2\pi i}{k_{z,1}} e^{ik_{z,1}(z+z^\prime)} \sum_{\alpha,\beta=\rm{s}, \rm{p}} r^{\alpha,\beta} \bm{\epsilon}_\alpha^+ \otimes \bm{\epsilon}_\beta^-~.
\end{equation}
Here and in what follows, we use the short notation $\bar{\bar{G}}_{\rm ref}(\bm{k},z,z^\prime)$ for the DGf tensor, where the double bar indicates the tensorial nature of the DGf and the $\omega$ argument is implicit. In Eq.~\eqref{eqn:app_ref_G} the symbol $\otimes$ refers to the tensor product in the real space. We, moreover, want to derive the duality relation between this expression and the reflected magnetic DGf $\bar{\bar{G}}_{\rm ref}^{\rm M}(\bm{k},z,z^\prime)$. 

We first introduce the $s-$ and $p-$ polarization vectors: 
\begin{equation}\label{eqn:s-polarization}
    \bm{\epsilon}_{s,n}^\pm \equiv \frac{\bm{k}\times\bm{z}}{k} = \frac{k_y \bm x - k_x\bm y}{k}~,
\end{equation}
\begin{equation}\label{eqn:p-polarization}
    \bm{\epsilon}_{p,n}^\pm \equiv \frac{\bm{\epsilon}_{s,n}^\pm\times\bm{k}_\pm}{|\bm{\epsilon}_{s,n}^\pm\times\bm{k}_\pm|} = \frac{k}{k_n}\bm z \mp \frac{k_{z,n}}{k_n}\frac{k_x \bm x + k_y\bm y}{k}~,
\end{equation}
where $\bm{k}_\pm \equiv k_x \bm{x} + k_y \bm y \pm k_{z,n} \bm z$ and $k_n = \frac{\omega}{c} \sqrt{\varepsilon_n}$, with $n = 1,2,3$ identifying the region of space, $n=1$ corresponding to $z>0$, $n=2$ corresponding to $-d<z<0$ and $n=3$ corresponding to $z<-d$. We also defined:
\begin{equation}
k_{z,n} =
\begin{cases}
    \sqrt{k_n^2 - k^2}, \qquad {\rm if} \quad k<k_n~,\\
    i\sqrt{k^2-k_n^2}, \qquad {\rm if} \quad k>k_n~.
\end{cases}
\end{equation}
Then, the free-space DGf is given by~\cite{app_peres}:
\begin{equation}
    \bar{\bar{G}}_{0}(\bm{k},z,z^\prime) = \frac{2\pi i}{k_{z,n}} e^{ik_{z,n}|z-z^\prime|}\left(\bm{\epsilon}_{s,n}^\pm \otimes \bm{\epsilon}_{s,n}^\pm + \bm{\epsilon}_{p,n}^\pm \otimes \bm{\epsilon}_{p,n}^\pm \right)~,
\end{equation}
where the $+$ sign is valid for $z>z^\prime$ and the $-$ sign is valid for $z<z^\prime$.
An electric dipole $\bm \mu(z_0)$ located at $z=z_0$ emits an electric field given by:
\begin{equation}
    \bm{E}_0(\bm{k},z) = \frac{\omega^2}{c^2} \bar{\bar{G}}_{0}(\bm{k},z,z_0) \cdot \bm{\mu}(z_0) = e^{ik_{z,n}|z-z_0|}\left(E^{0,\pm}_{s,n} \bm{\epsilon}_{s,n}^\pm + E^{0,\pm}_{p,n}\bm{\epsilon}_{p,n}^\pm \right)~,
\end{equation}
where we introduced the s- and p-polarization amplitudes:
\begin{equation}
    E^{0,\pm}_{s,n} =\frac{\omega^2}{c^2} \frac{2\pi i}{k_{z,n}}\bm{\epsilon}_{s,n}^\pm\cdot \bm{\mu}(z_0)~,
\end{equation}
\begin{equation}
    E^{0,\pm}_{p,n} =\frac{\omega^2}{c^2} \frac{2\pi i}{k_{z,n}}\bm{\epsilon}_{p,n}^\pm\cdot \bm{\mu}(z_0)~.
\end{equation}
Supposing that there is an interface located at $z=0$, the component of the emitted field evaluated at the interface itself, is given by the ``$-$'' propagating wave:
\begin{equation}
    \bm{E}(\bm{k},z) =  e^{ik_{z,n}z_0}\left(E^{0,-}_{s,n} \bm{\epsilon}_{s,n}^- + E^{0,-}_{p,n}\bm{\epsilon}_{p,n}^- \right)~.
\end{equation}
Since this electric field is already decomposed in $s-$ and $p-$ polarizations, we can obtain the reflected field by simply introducing the reflection coefficients (quotient between the reflected and incident electric field complex amplitudes \cite{appendix_Landau08}):
\begin{equation}
    E_{\alpha}^{\rm R} \equiv r^{\alpha,\beta}E_{\beta}^{\rm I}~, \quad \alpha,\beta={\rm s,p}~.
\end{equation}
The reflected field is then ($z_0>z>0$): 
\begin{equation}
\begin{split}
    \bm{E}_{\rm ref}(\bm{k},z) &=  e^{ik_{z,n}(z_0+z)}\left( E_{s}^{\rm R} \bm{\epsilon}_{s,n}^+ + E_{p}^{\rm R} \bm{\epsilon}_{p,n}^+\right)\\
    &=e^{ik_{z,n}(z_0+z)}\left( r^{s,s} E^{0,-}_{s,n} \bm{\epsilon}_{s,n}^+ + r^{p,s} E^{0,-}_{s,n} \bm{\epsilon}_{p,n}^+ + r^{s,p} E^{0,-}_{p,n} \bm{\epsilon}_{s,n}^+ + r^{p,p} E^{0,-}_{p,n}\bm{\epsilon}_{p,n}^+ \right)~.
\end{split}
\end{equation}
Making the electric field amplitudes explicit one obtains~\cite{app_peres}:
\begin{equation}
\begin{split}
    \bm{E}_{\rm ref}(\bm{k},z) &=  \frac{\omega^2}{c^2} \frac{2\pi i}{k_{z,n}} e^{ik_{z,n}(z_0+z)} \left( r^{s,s} \bm{\epsilon}_{s,n}^+ \otimes \bm{\epsilon}_{s,n}^- + r^{p,s}\bm{\epsilon}_{p,n}^+\otimes \bm{\epsilon}_{s,n}^- + r^{s,p} \bm{\epsilon}_{s,n}^+\otimes \bm{\epsilon}_{p,n}^- + r^{p,p} \bm{\epsilon}_{p,n}^+\otimes \bm{\epsilon}_{p,n}^- \right)\cdot \bm{\mu}(z_0)\\
    &= \frac{\omega^2}{c^2} \left[\frac{2\pi i}{k_{z,n}} e^{ik_{z,n}(z_0+z)} \sum_{\alpha,\beta\in\{{\rm s,p}\}} r^{\alpha,\beta} \bm{\epsilon}_{\alpha,n}^+ \otimes \bm{\epsilon}_{\beta,n}^-\right]\cdot\bm{\mu}(z_0)\\
    &\equiv \frac{\omega^2}{c^2} \bar{\bar{G}}_{\rm ref}(\bm{k},z,z_0)\cdot\bm{\mu}(z_0)~.
\end{split}
\end{equation}
The magnetic field that is scattered from the interface, coming from the electric dipole source, is obtained by using the Maxwell equation: $\bm{\nabla}\times\bm{E}(\bm{k},z) = i\omega/c \bm{B}$. We get:
\begin{equation}
\begin{split}
    \bm{B}_{\rm ref}(\bm{k},z) &= -i\frac{\omega}{c} \bm{\nabla}\times \bar{\bar{G}}_{\rm ref}(\bm{k},z,z_0)\cdot\bm{\mu}(z_0)~.
\end{split}
\end{equation}
Similar considerations can be made for the electromagnetic field generated by a magnetic moment $\bm{m}(z_0)$. In the absence of the interface, we have: 
\begin{equation}
    \bm{B}_0(\bm{k},z) = \frac{\omega^2}{c^2} \bar{\bar{G}}_{0}(\bm{k},z,z_0) \cdot \bm{m}(z_0) = e^{ik_{z,n}|z-z_0|}\left(B^{0,\pm}_{s,n} \bm{\epsilon}_{s,n}^\pm + B^{0,\pm}_{p,n}\bm{\epsilon}_{p,n}^\pm \right)~,
\end{equation}
where similarly to the electric field amplitudes we have introduced:
\begin{equation}
    B^{0,\pm}_{s,n} =\frac{\omega^2}{c^2} \frac{2\pi i}{k_{z,n}}\bm{\epsilon}_{s,n}^\pm\cdot \bm{m}(z_0)~,
\end{equation}
\begin{equation}
    B^{0,\pm}_{p,n} =\frac{\omega^2}{c^2} \frac{2\pi i}{k_{z,n}}\bm{\epsilon}_{p,n}^\pm\cdot \bm{m}(z_0)~.
\end{equation}
We can again evaluate the incoming field at the interface located at $z=0$ and derive the reflected magnetic field through the Fresnel coefficients. We introduce the following notation:
\begin{equation}
    B_{\alpha}^{\rm R} \equiv r^{\alpha,\beta}_{M}B_{\beta}^{\rm I}~, \quad \alpha,\beta=s,p~,
\end{equation}
and making the same considerations we have drawn for the electric field we obtain the following magnetic reflection DGf:
\begin{equation}
\bar{\bar{G}}^{\rm M}_{\rm ref}(\bm{k},z,z_0) = \frac{2\pi i}{k_{z,n}} e^{ik_{z,n}(z_0+z)} \sum_{\alpha,\beta\in\{{\rm s,p}\}} r_{M}^{i,j} \bm{\epsilon}_{\alpha,n}^+ \otimes \bm{\epsilon}_{\beta,n}^-~.
\end{equation}
Where we introduced the magnetic Fresnel coefficients $ r_{M}^{\alpha,\beta}$. Consider the $p$-polarized part of the incoming field:
\begin{equation}\label{app:incoming_magnetic}
    \bm{B}^I = B^I_p \bm{\epsilon}_{p,1}^+ e^{-ik_{z,1} z}e^{i(\bm{k}\cdot\bm{r} - \omega t )}~,
\end{equation}
this is accompanied by an electric field of the form:
\begin{equation}\label{app:incoming_electric}
    \bm{E}^I = -\frac{1}{\sqrt{\varepsilon_1}}B^I_p \bm{\epsilon}_{s,1} e^{-ik_{z,1} z}e^{i(\bm{k}\cdot\bm{r} - \omega t )}~.
\end{equation}
The reflected part is instead given by:
\begin{equation}\label{app:reflected_magnetic}
    \bm{B}^R = \left(B^R_s \bm{\epsilon}_{s,1} + B^R_p \bm{\epsilon}_{p,1}^-\right) e^{ik_{z,1} z}e^{i(\bm{k}\cdot\bm{r} - \omega t )}~,
\end{equation}
that gives a reflected electric field:
\begin{equation}\label{app:reflected_electric}
    \bm{E}^R = -\frac{1}{\sqrt{\varepsilon_1}}\left(B^R_p \bm{\epsilon}_{s,1} - B^R_s \bm{\epsilon}_{p,1}^-\right) e^{ik_{z,1} z}e^{i(\bm{k}\cdot\bm{r} - \omega t )}~.
\end{equation}
We conclude that calling the amplitudes of the polarization vectors of the electric fields as:
\begin{equation}
    E^I_s = -\frac{1}{\sqrt{\varepsilon_1}}B^I_p,\quad E^R_s = -\frac{1}{\sqrt{\varepsilon_1}}B^R_p,\quad E^R_p = \frac{1}{\sqrt{\varepsilon_1}}B^R_s~,
\end{equation}
the reflection coefficients of the magnetic field are related to the electric ones by:
\begin{equation}
    r^{p,p}_{M} = \frac{B_{p}^{\rm R}}{B_{p}^{\rm I}} = \frac{E_{s}^{\rm R}}{E_{s}^{\rm I}} =  r^{s,s}~,
\end{equation}
\begin{equation}
    r^{s,p}_{M} = \frac{B_{s}^{\rm R}}{B_{p}^{\rm I}} = -\frac{E_{p}^{\rm R}}{E_{s}^{\rm I}} =  -r^{p,s}~.
\end{equation}
A similar calculation for the $s-$ polarization of the incident field gives the following:
\begin{equation}
    r^{s,s}_{M} = \frac{B_{s}^{\rm R}}{B_{s}^{\rm I}} = \frac{E_{p}^{\rm R}}{E_{p}^{\rm I}} =  r^{p,p}~,
\end{equation}
\begin{equation}
    r^{p,s}_{M} = \frac{B_{p}^{\rm R}}{B_{s}^{\rm I}} = -\frac{E_{s}^{\rm R}}{E_{p}^{\rm I}} =  -r^{s,p}~.
\end{equation}
Finally, the magnetic reflection tensor is given by:
\begin{equation}
    r_{M} = \left(\begin{array}{cc}
        r_{M}^{s,s} & r_{M}^{s,p} \\
        r_{M}^{p,s} & r_{M}^{p,p}
    \end{array}\right) = \left(\begin{array}{cc}
        r^{p,p} & -r^{p,s} \\
        -r^{s,p} & r^{s,s}
    \end{array}\right)~.
\end{equation}
The scattering Green's function is then obtained, for the magnetic moment, as:
\begin{equation}\label{eqn:magnetic_DGf_ref}
\bar{\bar{G}}_{\rm ref}^{\rm M}(\bm{k},z,z_0) = \frac{4\pi i}{2k_{z,n}} e^{ik_{z,n}(z_0+z)} \sum_{\alpha,\beta\in\{{\rm s,p}\}} (-1)^{1-\delta_{\alpha\beta}} r^{\bar{\alpha},\bar{\beta}} \bm{\epsilon}_{\alpha,n}^+ \otimes \bm{\epsilon}_{\beta,n}^-~,
\end{equation}
where $(-1)^{1-\delta_{\alpha\beta}} = 1$ if $\alpha=\beta$ and $(-1)^{1-\delta_{\alpha\beta}} = -1$ when $\alpha\neq \beta$, where the index $\bar{\xi} = p,s$ when $\xi=s,p$. Eq.~\eqref{eqn:magnetic_DGf_ref} is the main result of this appendix. It shows how one should construct the reflection magnetic DGf from the reflection tensor associated to the electric field. In particular one should inter-change $s-$ and $p-$ polarization,~e.g. $r^{s,s}\to r^{p,p}$, and change the sign of the off diagonal elements,~e.g. $r^{s,p}\to-r^{p,s}$.
\subsection{Symmetries of $r^{\alpha,\beta}$}\label{appendixC1}
By means of the reciprocity principle~\cite{app_Tai1994, appendix_Landau08}, the (electric) scattering DGf satisfies:
\begin{equation}
    \bar{\bar{G}}_{\rm ref}(\bm{k},z,z^\prime) = \left[\bar{\bar{G}}_{\rm ref}(-\bm{k},z^\prime,z)\right]^{\rm T}~.
\end{equation}
This condition directly implies that:
\begin{equation}\label{eqn:reciprocity}
    \sum_{\alpha,\beta}r^{\alpha,\beta}\left[\bm{\epsilon}_\alpha^+(\bm k) \otimes \bm{\epsilon}_\beta^-(\bm k)\right] = \sum_{\alpha,\beta}r^{\alpha,\beta}\left[\bm{\epsilon}_\alpha^+(-\bm k) \otimes \bm{\epsilon}_\beta^-(-\bm k)\right]^{\rm T}~.
\end{equation}
Exploiting the tensor product between the polarization vectors (this calculation is presented below in Appendix~\ref{appendixD}, Eqs.~\eqref{eqn:tensor_prod1}-\eqref{eqn:tensor_prod4}), one can easily show that the reciprocity brings to:
\begin{equation}
    r^{s,p} = -r^{p,s}~.
\end{equation}
Despite this symmetry constraint is valid for every reciprocal system, due to the definitions~\eqref{eqn:s-polarization} and~\eqref{eqn:s-polarization} of $\bm{\epsilon}_{s}^\pm$ and $\bm{\epsilon}_{p}^\pm$, we see that in order to have $r^{s,p}$ and $r^{p,s}$ different from zero, one should break inversion with respect to a point and, at least, a plane, or break time-reversal symmetry. The former condition is exactly the condition for chirality~\cite{flack}. The latter condition is obtained when a field $\bm{\mathcal{M}}$ that is odd under time reversal is present (an external magnetic field applied at the interface, a magnetic ordered substrate, etc.). In this case, the right-hand side of Eq.~\eqref{eqn:reciprocity} should be intended with that field inverted. One can then generalize the relations between the elements of the reflection tensor $r^{\alpha,\beta}$ as:
\begin{gather}
    r^{s,s}(\bm{\mathcal{M}}) = r^{s,s}(-\bm{\mathcal{M}})~,\\
    r^{p,p}(\bm{\mathcal{M}}) = r^{p,p}(-\bm{\mathcal{M}})~,\\
    r^{s,p}(\bm{\mathcal{M}}) = -r^{p,s}(-\bm{\mathcal{M}})~,\\
    r^{p,s}(\bm{\mathcal{M}}) = -r^{s,p}(-\bm{\mathcal{M}})~.
\end{gather}
Thanks to these equations, one can in principle obtain a symmetric reflection tensor $r^{\alpha,\beta}$,~i.e. $r^{s,p}=r^{p,s}\neq0$. This is the case, for example, of a single layer graphene interface under the influence of an external magnetic field~\cite{app_kort-kamp_prb_2015}. 
\section{Full calculation of the dissipated power for vertically and arbitrarily oriented chiral dipole}\label{appendixD}
As we have seen in the main text, the environment contribution to the dissipated power is composed by four contributions:
\begin{gather}
    P_{e,e}^{\rm ref}(\omega) = \frac{\omega^3}{2c^2}\int\frac{d^2\bm{k}}{(2\pi)^2}\Im \left[\bm{\mu}\cdot  \bar{\bar{G}}_{{\rm ref}}(\bm{k},z_0,z_0,\omega) \cdot\bm{\mu}\right]~,\label{eqn:app_ref_dissipated_power1} \\
    P_{e,m}^{\rm ref}(\omega) = \pm \frac{\omega^2}{2c}\int\frac{d^2\bm{k}}{(2\pi)^2}\Im \left[\bm{\mu}\cdot \bm{\nabla}\times \bar{\bar{G}}_{{\rm ref}}^{\rm M}(\bm{k},z_0,z_0,\omega) \cdot\bm{m}\right]~,\label{eqn:app_ref_dissipated_power2} \\
    P_{m,e}^{\rm ref}(\omega) = \pm\frac{\omega^2}{2c}\int\frac{d^2\bm{k}}{(2\pi)^2}\Im \left[\bm{m}\cdot\bm{\nabla}\times \bar{\bar{G}}_{{\rm ref}}(\bm{k},z_0,z_0,\omega) \cdot \bm{\mu}\right]\label{eqn:app_ref_dissipated_power3} ~,\\
    P_{m,m}^{\rm ref}(\omega) = \frac{\omega^3}{2c^3}\int\frac{d^2\bm{k}}{(2\pi)^2}\Im \left[\bm{m}\cdot \bar{\bar{G}}_{{\rm ref}}^{\rm M}(\bm{k},z_0,z_0,\omega) \cdot\bm{m}\right]\label{eqn:app_ref_dissipated_power4}~.
\end{gather}
with the reflection DGfs given by the reflection coefficients of the interface (Eq.~\eqref{eqn:app_ref_G} and~\eqref{eqn:magnetic_DGf_ref}):
\begin{equation}\label{eqn:Gsc_def2}
    \bar{\bar{G}}_{\rm ref}(\bm{k},z,z^\prime) = \frac{4\pi i}{2k_{z,1}} e^{ik_{z,1}(z+z^\prime)} \sum_{\alpha,\beta=\rm{s}, \rm{p}} r^{\alpha,\beta} \bm{\epsilon}_\alpha^+ \otimes \bm{\epsilon}_\beta^-~,
\end{equation}
\begin{equation}\label{eqn:GscM_def2}
\bar{\bar{G}}_{\rm ref}^{\rm M}(\bm{k},z,z^\prime) = \frac{4\pi i}{2k_{z,n}} e^{ik_{z,n}(z^\prime+z)} \sum_{\alpha,\beta\in\{{\rm s,p}\}} (-1)^{1-\delta_{\alpha\beta}} r^{\bar{\alpha},\bar{\beta}} \bm{\epsilon}_{\alpha,n}^+ \otimes \bm{\epsilon}_{\beta,n}^-~.
\end{equation}

From the definitions of the $s-$ and $p-$ polarization vectors (Eq.~\eqref{eqn:s-polarization} and Eq.~\eqref{eqn:p-polarization} respectively), we can make the tensor product $\bm{\epsilon}_\alpha^+ \otimes \bm{\epsilon}_\beta^-$ explicit for every combination $\alpha,\beta\in\{s,p\}$~\cite{app_peres} (we consider the region above the interface $z>0$, as we are dealing with reflection):
\begin{equation}\label{eqn:tensor_prod1}
    \bm{\epsilon}_{s,1}^+ \otimes \bm{\epsilon}_{s,1}^- = \frac{1}{k^2}\left[\begin{array}{ccc}
        k_y^2 & -k_xk_y & 0 \\
        -k_xk_y & k_x^2 & 0 \\
        0 & 0 & 0 \\
    \end{array}\right]~,
\end{equation}
\begin{equation}\label{eqn:tensor_prod2}
    \bm{\epsilon}_{p,1}^+ \otimes \bm{\epsilon}_{p,1}^- = \frac{1}{k_1^2}\left[\begin{array}{ccc}
        -k_x^2k_{z,1}^2/k^2 & -k_xk_yk_{z,1}^2/k^2 & -k_x k_{z,1} \\
        -k_xk_yk_{z,1}^2/k^2 & -k_y^2k_{z,1}^2/k^2 & -k_y k_{z,1} \\
        k_x k_{z,1} & k_y k_{z,1} & k^2
    \end{array}\right]~,
\end{equation}
\begin{equation}\label{eqn:tensor_prod3}
    \bm{\epsilon}_{s,1}^+ \otimes \bm{\epsilon}_{p,1}^- = \frac{1}{k_1}\left[\begin{array}{ccc}
        k_xk_yk_{z,1}/k^2 & k_y^2 k_{z,1}/k^2 & k_y \\
        -k_x^2 k_{z,1}/k^2 & -k_xk_yk_{z,1}/k^2 & -k_x \\
        0 & 0 & 0 \\
    \end{array}\right]~,
\end{equation}
\begin{equation}\label{eqn:tensor_prod4}
    \bm{\epsilon}_{p,1}^+ \otimes \bm{\epsilon}_{s,1}^- = \frac{1}{k_1}\left[\begin{array}{ccc}
        -k_xk_yk_{z,1}/k^2 & k_x^2 k_{z,1}/k^2 & 0 \\
        -k_y^2 k_{z,1}/k^2 & k_xk_yk_{z,1}/k^2 & 0 \\
        k_y & -k_x & 0 \\
    \end{array}\right]~.
\end{equation}
Here we explicit also the curl of the reflected Green's function:
\begin{equation}
    \bm{\nabla}\times\bar{\bar{G}}_{\rm ref}(\bm{k},z,z^\prime) =   \left(-ik_x, -ik_y, \partial_z\right)\times\frac{i}{2k_{z,1}} e^{ik_{z,1}(z+z^\prime)} \sum_{\alpha,\beta=\rm{s}, \rm{p}} r^{i,j} \bm{\epsilon}_\alpha^+ \otimes \bm{\epsilon}_\beta^-~.
\end{equation}
For the sake of clarity we make the following observation:
\begin{equation}
    \left[\bm{\nabla} \times \bar{\bar{T}}\right]_{ij} = \epsilon_{ik\ell} \partial_k T_{\ell j}~,
\end{equation}
and when the tensor $\bar{\bar{T}}$ is expressed as a tensor product between two vectors $\bar{\bar{T}} = \bm{v}\otimes\bm{w}$:
\begin{equation}
    \left[\bm{\nabla} \times \bar{\bar{T}}\right]_{ij} = \epsilon_{ik\ell} \partial_k T_{\ell j} = \epsilon_{ik\ell} \partial_k v_{\ell} w_{j} = \left[\left(\bm{\nabla} \times \bm{v}\right)\otimes \bm{w}\right]_{ij}~.
\end{equation}
Proceeding in order, we start from $i = s$:
\begin{equation}
    \left(-ik_x,  -ik_y, \partial_z\right)\times\left[\frac{i}{2k_{z,1}} e^{ik_{z,1}(z+z^\prime)} \bm{\epsilon}_s^+\right]= -\frac{1}{2}\frac{k_1}{k_{z,1}}e^{ik_{z,1}(z+z^\prime)}\bm{\epsilon}_{p}^{-}\equiv e^{ik_{z,1}(z+z^\prime)}\tilde{\bm{\epsilon}}_{p}^{-}~,
\end{equation}
while for $i = p$ we get:
\begin{equation}
    \left(-ik_x,  -ik_y, \partial_z\right)\times\left[\frac{i}{2k_{z,1}} e^{ik_{z,1}(z+z^\prime)} \bm{\epsilon}_p^+\right]= \frac{k^2 - k_{z,1}^2}{2k_{z,1}k_1}e^{ik_{z,1}(z+z^\prime)}\bm{\epsilon}_{s}^{-}\equiv e^{ik_{z,1}(z+z^\prime)}\tilde{\bm{\epsilon}}_{s}^{-}~.
\end{equation}
We conclude that the curl applied to $\bar{\bar{G}}_{\rm ref}(\bm{k},z,z^\prime)$ gives:
\begin{equation}\label{eqn:curl_green_scattered}
    \bm{\nabla} \times\bar{\bar{G}}_{\rm ref}(\bm{k},z,z^\prime) = 4\pi e^{ik_{z,1}(z+z^\prime)} \sum_{\alpha,\beta=\rm{s}, \rm{p}} r^{\alpha,\beta} \tilde{\bm{\epsilon}}_{\bar \alpha}^- \otimes \bm{\epsilon}_\beta^-~,
\end{equation}
where the index $\bar{\alpha} = p,s$ when $\alpha=s,p$ and the modified polarization vectors are defined by:
\begin{equation}
   \tilde{\bm{\epsilon}}_{p}^{-}\equiv -\frac{1}{2}\frac{k_1}{k_{z,1}}\bm{\epsilon}_{p}^{-}~,
\end{equation}
\begin{equation}
    \tilde{\bm{\epsilon}}_{s}^{-}\equiv \frac{k^2 - k_{z,1}^2}{2k_{z,1}k_1}\bm{\epsilon}_{s}^{-}~.
\end{equation}
A similar derivation for the magnetic DGf gives:
\begin{equation}\label{eqn:curl_green_scattered_mangetic_moment}
    \bm{\nabla} \times\bar{\bar{G}}_{\rm ref}^{\rm M}(\bm{k},z,z^\prime) = 4\pi  e^{ik_{z,1}(z+z^\prime)} \sum_{\alpha,\beta=\rm{s}, \rm{p}} (-1)^{1-\delta_{\alpha\beta}} r^{\bar{\alpha},\bar{\beta}} \tilde{\bm{\epsilon}}_{\bar i}^- \otimes \bm{\epsilon}_j^-~.
\end{equation}
In order to use equation~\eqref{eqn:curl_green_scattered}, we must exploit the tensor product between the modified polarization vectors and the standard ones~\eqref{eqn:s-polarization} and~\eqref{eqn:p-polarization}. We have:
\begin{equation}\label{eqn:tensor_product_curl1}
    \tilde{\bm{\epsilon}}_{s,1}^- \otimes \bm{\epsilon}_{s,1}^- = \frac{k^2 - k_{z,1}^2}{2k_{z,1}k_1}\frac{1}{k^2}\left[\begin{array}{ccc}
        k_y^2 & -k_xk_y & 0 \\
        -k_xk_y & k_x^2 & 0 \\
        0 & 0 & 0 \\
    \end{array}\right]~,
\end{equation}
\begin{equation}\label{eqn:tensor_product_curl2}
    \tilde{\bm{\epsilon}}_{p,1}^- \otimes \bm{\epsilon}_{p,1}^- = -\frac{1}{2}\frac{1}{k_{z,1}k_1}\left[\begin{array}{ccc}
        k_x^2k_{z,1}^2/k^2 & k_xk_yk_{z,1}^2/k^2 & k_x k_{z,1} \\
        k_xk_yk_{z,1}^2/k^2 & k_y^2k_{z,1}^2/k^2 & k_y k_{z,1} \\
        k_x k_{z,1} & k_y k_{z,1} & k^2
    \end{array}\right]~,
\end{equation}
\begin{equation}\label{eqn:tensor_product_curl3}
    \tilde{\bm{\epsilon}}_{s,1}^- \otimes \bm{\epsilon}_{p,1}^- = \frac{k^2 - k_{z,1}^2}{2k_{z,1}k_1^2} \left[\begin{array}{ccc}
        k_xk_yk_{z,1}/k^2 & k_y^2 k_{z,1}/k^2 & k_y \\
        -k_x^2 k_{z,1}/k^2 & -k_xk_yk_{z,1}/k^2 & -k_x \\
        0 & 0 & 0 \\
    \end{array}\right]~,
\end{equation}
\begin{equation}\label{eqn:tensor_product_curl4}
    \tilde{\bm{\epsilon}}_{p,1}^- \otimes \bm{\epsilon}_{s,1}^- = -\frac{1}{2k_{z,1}} \left[\begin{array}{ccc}
        k_xk_yk_{z,1}/k^2 & -k_x^2 k_{z,1}/k^2 & 0 \\
        k_y^2 k_{z,1}/k^2 & -k_xk_yk_{z,1}/k^2 & 0 \\
        k_y & -k_x & 0 \\
    \end{array}\right]~.
\end{equation}
\subsection{Vertical orientation}
Assuming a chiral dipole $\bm{d}_{\pm}$ that consists of a vertical electric dipole and an inclined magnetic moment: $\bm{d}_{\pm} = \left\{(\pm i)\mu_0\hat{\bm z} + m_0[\hat{\bm x}\sin\varphi + \hat{\bm z}\cos\varphi]\right\}e^{-i\omega t}$ we can, without lost of generality, fix the coordinate system such that the in-plane momentum $\bm{k}$ lies along the $x$ direction. The polarization vectors are then given by:
\begin{gather}
    \bm{\epsilon}_{s,n}^{\pm} = \bm y~,\\
    \bm{\epsilon}_{p,n}^{\pm} = \frac{\mp k_{z,n}\bm x + k\bm z}{k_n}~.
\end{gather}
The diagonal dissipated power contributions can be obtained with the following calculation:
\begin{equation}
\begin{split}
    P_{e,e}^{\rm ref} &= \frac{\omega^3}{2c^2}\int\frac{d^2\bm{k}}{(2\pi)^2}\Im \left[\bm{\mu}\cdot\bar{\bar{G}}_{{\rm ref}}(\bm{k},z_0,z_0,\omega)\cdot \bm{\mu}\right]=\\
    &= \frac{\omega^3}{2 c^2}\mu_0^2 \sum_{\alpha,\beta=\rm{s}, \rm{p}}\int\frac{d^2\bm{k}}{(2\pi)^2}\Im \left[\frac{4\pi i}{2k_{z,1}} e^{2ik_{z,1} z_0}r^{\alpha,\beta}\bm{z} \cdot  \bm{\epsilon}_\alpha^+ \otimes \bm{\epsilon}_\beta^-\cdot \bm{z}\right]=\\
    &= \frac{\omega^3}{2 c^2}\mu_0^2 \int\frac{d^2\bm{k}}{(2\pi)^2}\frac{k^2}{k_1^2}\Im \left[\frac{4\pi i}{2k_{z,1}} e^{2ik_{z,1} z_0}r^{p,p}\right]~,
\end{split}
\end{equation}
similarly for the magnetic part we have:
\begin{equation}
\begin{split}
    P_{m,m}^{\rm ref} &= \frac{\omega^3}{2c^2}\int\frac{d^2\bm{k}}{(2\pi)^2}\Im \left[\bm{m}^*\cdot \bar{\bar{G}}_{{\rm ref}}^{\rm M}(\bm{k},z_0,z_0,\omega)\cdot \bm{m}\right] \\
    &= \frac{\omega^3}{2 c^2}\mu_0^2 \sum_{\alpha,\beta=\rm{s}, \rm{p}}(-1)^{1-\delta_{\alpha\beta}}\int\frac{d^2\bm{k}}{(2\pi)^2}\Im \left[\frac{4\pi i}{2k_{z,1}} e^{2ik_{z,1} z_0} r^{\bar{\alpha},\bar{\beta}} \left(\bm{z} \cdot  \bm{\epsilon}_\alpha^+ \otimes \bm{\epsilon}_\beta^-\cdot \bm{z}\cos^2\varphi + \bm{x} \cdot  \bm{\epsilon}_\alpha^+ \otimes \bm{\epsilon}_\beta^-\cdot \bm{x}\sin^2\varphi\right)\right]\\
    &{\quad}+ \frac{\omega^3}{2 c^2}\mu_0^2 \sum_{\alpha,\beta=\rm{s}, \rm{p}}(-1)^{1-\delta_{\alpha\beta}}\int\frac{d^2\bm{k}}{(2\pi)^2}\Im \left[\frac{4\pi i}{2k_{z,1}} e^{2ik_{z,1} z_0}r^{\bar{\alpha},\bar{\beta}}\left(\bm{x} \cdot  \bm{\epsilon}_\alpha^+ \otimes \bm{\epsilon}_\beta^-\cdot \bm{z} + \bm{z} \cdot  \bm{\epsilon}_\alpha^+ \otimes \bm{\epsilon}_\beta^-\cdot \bm{x}\right)\sin\varphi\cos\varphi\right]\\
    &= \frac{\omega^3}{2 c^2}\mu_0^2 \sum_{\alpha,\beta=\rm{s}, \rm{p}}\int\frac{d^2\bm{k}}{(2\pi)^2}\Im \left[\frac{4\pi i}{2k_{z,1}k_1^2} e^{2ik_{z,1} z_0}r^{s,s}\left(k^2\cos^2\varphi -k_{z,1}^2\sin^2\varphi\right)\right]~.
\end{split}
\end{equation}
The cross contributions are instead obtained as follows:
\begin{equation}
\begin{split}
    P_{e,m}^{\rm ref} &= \pm\frac{\omega^2}{2 c}\int\frac{d^2\bm{k}}{(2\pi)^2}\Im \left[\bm{\mu}\cdot \left.\bm{\nabla}\times\bar{\bar{G}}_{\rm ref}^{\rm M}(\bm{k},z,z_0,\omega)\right|_{z=z_0} \cdot \bm{m}\right] \\
    &= \pm \frac{\omega^2}{2c}\mu_0 m_0\sum_{\alpha,\beta=\rm{s}, \rm{p}}(-1)^{1-\delta_{\alpha\beta}}\int\frac{d^2\bm{k}}{(2\pi)^2}\Im \left[ 4\pi  e^{2ik_{z,1}z_0} r^{\bar{\alpha},\bar{\beta}}  \bm{z}\cdot \tilde{\bm{\epsilon}}_{\alpha,1}^- \otimes \bm{\epsilon}_{\beta,1}^- \cdot \bm{z}\cos\varphi\right]\\
    &{\quad}\pm \frac{\omega^2}{2c}\mu_0 m_0\sum_{\alpha,\beta=\rm{s}, \rm{p}}(-1)^{1-\delta_{\alpha\beta}}\int\frac{d^2\bm{k}}{(2\pi)^2}\Im \left[ 4\pi  e^{2ik_{z,1}z_0} r^{\bar{\alpha},\bar{\beta}}\bm{z}\cdot \tilde{\bm{\epsilon}}_{\alpha,1}^- \otimes \bm{\epsilon}_{\beta,1}^- \cdot \bm{x}\sin\varphi\right]\\
    &= \mp \frac{\omega^2}{2c}\mu_0 m_0\int\frac{d^2\bm{k}}{(2\pi)^2}\Im \left[ 4\pi  e^{2ik_{z,1}z_0} r^{p,s}\left(-\frac{1}{2}\frac{k^2}{k_{z,1}k_1}\cos\varphi - \frac{k}{2k_1} \sin\varphi\right)\right]\\
    &=\pm \frac{\omega^2}{4c}\mu_0 m_0\int\frac{d^2\bm{k}}{(2\pi)^2} \Im\left[\frac{4\pi k}{k_1 k_{z,1}}e^{2 i k_{z,1} z_0} r^{p,s} (k \cos \varphi +k_{z,1} \sin\varphi)\right]~,
\end{split}
\end{equation}
similarly:
\begin{equation}\label{eqn:app_Pme}
\begin{split}
    P_{m,e}^{\rm ref} &= \pm\frac{\omega^2}{2c}\int\frac{d^2\bm{k}}{(2\pi)^2}\Im \left[\bm{m}\cdot \left.\bm{\nabla}\times\bar{\bar{G}}_{\rm sc}(\bm{k},z,z_0,\omega)\right|_{z=z_0} \cdot \bm{\mu}\right]\\
    &= \mp \frac{\omega^2}{4c}\mu_0 m_0\int\frac{d^2\bm{k}}{(2\pi)^2} \Im\left[\frac{4\pi  k}{k_1 k_{z,1}}e^{2 i k_{z,1} z_0} r^{s,p} (k \cos \varphi +k_{z,1} \sin\varphi)\right]~.
\end{split}
\end{equation}
\subsection{Arbitrary orientation}
In order to describe an arbitrarily oriented dipole, we introduce the polar angle $\phi$ and the azimuth angle $\vartheta$ :
\begin{equation}
    \bm{d}_{\pm} = \pm i\mu_0\left(\sin\vartheta\cos\phi, \sin\vartheta\sin\phi ,\cos\vartheta \right) + m_0\left(\sin\vartheta\cos\phi, \sin\vartheta\sin\phi ,\cos\vartheta \right)~,
\end{equation}
where $\mu_0$ and $m_0$ are the modulus of the electric and magnetic dipoles. Substituting this expression in Eqs.~\eqref{eqn:app_ref_dissipated_power1}--\eqref{eqn:app_ref_dissipated_power4} and making use of the full expressions for the reflection contributions of the DGfs~\eqref{eqn:Gsc_def2} and~\eqref{eqn:GscM_def2} and their curl~\eqref{eqn:curl_green_scattered} and~\eqref{eqn:curl_green_scattered_mangetic_moment}, we explicitly get:
\begin{equation}
\begin{split}
    P_{e,e}^{\rm ref}=& \pi \frac{\omega ^3}{c^2} \mu_0^2\int\frac{d^2\bm{k}}{(2\pi)^2}  \left\{ \Re\left[\frac{e^{2 i k_{z,1} z_0}}{k_1^2 k_{z,1}} r^{p,p} k^2\cos^2\vartheta\right] +\Re\left[ \frac{e^{2 i k_{z,1} z_0}}{k_1^2 k_{z,1}}k\sin\vartheta\cos\vartheta(r^{p,p} k_{z,1} \cos\phi - r^{p,s} k_1 \sin\phi)\right]\right.\\
    &\left.- \sin \vartheta\cos\phi \Re\left[\frac{e^{2 i k_{z,1} z_0}}{k_1^2} (r^{p,p} k \cos \vartheta - r^{p,s} k_1 \sin \phi  \sin \vartheta + r^{p,p} k_{z,1}\cos \phi \sin \vartheta )\right]\right.\\
    &\left.-\sin \vartheta \sin \phi  \Re\left[\frac{e^{2 i k_{z,1} z_0}}{k_1 k_{z,1}} ( r^{s,p} k\cos\vartheta -  r^{s,s} k_1 \sin\phi \sin\vartheta +  r^{s,p} k_{z,1} \cos \phi  \sin \vartheta )\right]\right\}~,
\end{split}
\end{equation}
\begin{equation}
\begin{split}
    P_{m,m}^{\rm ref}=& \pi \frac{\omega ^3}{c^2} m_0^2\int\frac{d^2\bm{k}}{(2\pi)^2}  \left\{ \Re\left[\frac{e^{2 i k_{z,1} z_0}}{k_1^2 k_{z,1}} r^{s,s} k^2\cos^2\vartheta\right] +\Re\left[ \frac{e^{2 i k_{z,1} z_0}}{k_1^2 k_{z,1}}k\sin\vartheta\cos\vartheta(r^{s,s} k_{z,1} \cos\phi + r^{s,p} k_1 \sin\phi)\right]\right.\\
    &\left.- \sin \vartheta\cos\phi \Re\left[\frac{e^{2 i k_{z,1} z_0}}{k_1^2} (r^{s,s} k \cos \vartheta + r^{s,p} k_1 \sin \phi  \sin \vartheta + r^{s,s} k_{z,1}\cos \phi \sin \vartheta )\right]\right.\\
    &\left.+\sin \vartheta \sin \phi  \Re\left[\frac{e^{2 i k_{z,1} z_0}}{k_1 k_{z,1}} ( r^{p,s} k\cos\vartheta +  r^{p,p} k_1 \sin\phi \sin\vartheta +  r^{p,s} k_{z,1} \cos \phi  \sin \vartheta )\right]\right\}~.
\end{split}
\end{equation}
\begin{equation}
    \begin{split}
        P_{e,m}^{\rm ref} = &\pm \pi \frac{\omega ^2}{c} \mu_0 m_0 \int\frac{d^2\bm{k}}{(2\pi)^2}  \sin \phi \sin \vartheta \Im\left[\frac{e^{2 i k_{z,1} z_{0}}}{k_{1}^2 k_{z,1}}\left(k_{z,1}^2-k^2\right) (r^{s,s}  k_{z,1}\cos \phi \sin \vartheta + r^{s,s}   k\cos \vartheta + r^{s,p}  k_1\sin \phi \sin \vartheta)\right]\\
        &\pm \frac{\omega ^2}{4 c}\mu_0 m_0\int\frac{d^2\bm{k}}{(2\pi)^2}  \Im\left[\frac{e^{2 i k_{z,1} z_{0}}}{k_{1}k_{z,1}} \left(k_{z,1}\cos \phi  \sin \vartheta  + k\cos \vartheta \right)  ( r^{p,s} k \cos \vartheta+  r^{p,p} k_{1} \sin \phi \sin \vartheta + r^{p,s} k_{z,1} \cos \phi \sin \vartheta )\right]~,
    \end{split}
\end{equation}
\begin{equation}
    \begin{split}
        P_{m,e}^{\rm ref} = &\pm \pi \frac{\omega ^2}{c} \mu_0 m_0 \int\frac{d^2\bm{k}}{(2\pi)^2}  \sin \phi \sin \vartheta \Im\left[\frac{e^{2 i k_{z,1} z_{0}}}{k_{1}^2 k_{z,1}}\left(k_{z,1}^2-k^2\right) (r^{p,p}  k_{z,1}\cos \phi \sin \vartheta + r^{p,p}   k\cos \vartheta - r^{p,s}  k_1\sin \phi \sin \vartheta)\right]\\
        &\mp \frac{\omega ^2}{4 c}\mu_0 m_0\int\frac{d^2\bm{k}}{(2\pi)^2}  \Im\left[\frac{e^{2 i k_{z,1} z_{0}}}{k_{1}k_{z,1}} \left(k_{z,1}\cos \phi  \sin \vartheta  + k\cos \vartheta \right)  ( r^{s,p} k \cos \vartheta - r^{s,s} k_{1} \sin \phi \sin \vartheta  + r^{s,p} k_{z,1} \cos \phi \sin \vartheta )\right]~.
    \end{split}
\end{equation}
From these results, the vertical orientation is obtained by setting $\vartheta = 0$. Similarly, for a horizontally oriented chiral dipole $\bm{d}_{\pm}$, the dissipated powers can be obtained by using $\vartheta = \pi/2$, leaving $\phi\in[0, 2\pi)$.
\section{Solution of the electromagnetic scattering problem on a twisted bilayer interface: derivation of the reflection tensor}\label{appendixE}
We can solve the scattering problem by separating the cases of $s-$ and $p-$ polarization~\cite{app_novotny, app_kort-kamp_prb_2015}. We report the steps for the first one, with the intention that for the $p-$ polarized incident wave the method is exactly the same. For an $s-$ polarized incoming wave we have~\cite{app_kort-kamp_prb_2015}:
\begin{equation}\label{eqn:incoming_electric}
    \bm{E}_I = E_I^s \bm{\epsilon}_{\rm s,1}^{+} e^{-ik_{z,1} z}e^{i(\bm{k}\cdot\bm{r} - \omega t )}~,
\end{equation}
where $\bm{\epsilon}_{{\rm s},n}^{\pm}$ was defined in Eq.~\eqref{eqn:s-polarization}, with $n$ labelling the region of space described by the dielectric constant $\varepsilon_n$. The magnetic field associated to this electric field is:
\begin{equation}\label{eqn:incoming_magnetic}
    \bm{B}_I = -\sqrt{\varepsilon_1}E_I^s \bm{\epsilon}_{\rm p,1}^+ e^{-ik_{z,1} z}e^{i(\bm{k}\cdot\bm{r} - \omega t )}~,
\end{equation}
where $\bm{\epsilon}_{{\rm p},n}^\pm$ was introduced in Eq.~\eqref{eqn:p-polarization}. This incident wave will produce the reflected EM field $\bm{E}_R$, $\bm{B}_R$ and the transmitted one $\bm{E}_{T}$, $\bm{B}_{T}$ in the bottom dielectric region:
\begin{equation}\label{eqn:E_reflected}
    \bm{E}_R = \left(E_R^s \bm{\epsilon}_{\rm s,1} + E_R^p \bm{\epsilon}_{\rm p,1}^-\right) e^{ik_{z,1} z}e^{i(\bm{k}\cdot\bm{r} - \omega t )}~,
\end{equation}
\begin{equation}\label{eqn:B_reflected}
    \bm{B}_R = \sqrt{\varepsilon_1}\left(E_R^p \bm{\epsilon}_{\rm s,1} - E_R^s \bm{\epsilon}_{\rm p,1}^-\right) e^{ik_{z,1} z}e^{i(\bm{k}\cdot\bm{r} - \omega t )}~,
\end{equation}
\begin{equation}
    \bm{E}_{in} = \left[\left(E_{in}^{s,-} \bm{\epsilon}_{\rm s,2} + E_{in}^{p,-} \bm{\epsilon}_{\rm p,2}^-\right) e^{ik_{z,2} z} + \left(E_{in}^{s,+} \bm{\epsilon}_{\rm s,2} + E_{in}^{p,+} \bm{\epsilon}_{\rm p,2}^+\right) e^{-ik_{z,2} z}\right]e^{i(\bm{k}\cdot\bm{r} - \omega t )}~,
\end{equation}
\begin{equation}
    \bm{B}_{in} =  \sqrt{\varepsilon_2}\left[\left(E_{in}^{p,-}\bm{\epsilon}_{\rm s,2}-E_{in}^{s,-} \bm{\epsilon}_{\rm p,2}^-  \right) e^{ik_{z,2} z} + \left(E_{in}^{p,+}\bm{\epsilon}_{\rm s,2} - E_{in}^{s,+} \bm{\epsilon}_{\rm p,2}^+  \right) e^{-ik_{z,2} z}\right]e^{i(\bm{k}\cdot\bm{r} - \omega t )}~,
\end{equation}
\begin{equation}
    \bm{E}_T = \left(E_T^s \bm{\epsilon}_{\rm s,3} + E_T^p \bm{\epsilon}_{\rm p,3}^+\right) e^{-ik_{z,3} z}e^{i(\bm{k}\cdot\bm{r} - \omega t )}~.
\end{equation}
\begin{equation}
    \bm{B}_T =  \sqrt{\varepsilon_3}\left(E_T^p \bm{\epsilon}_{\rm s,3} - E_T^s \bm{\epsilon}_{\rm p,3}^+\right) e^{-ik_{z,3} z}e^{i(\bm{k}\cdot\bm{r} - \omega t )}~,
\end{equation}
We have to relate the unknowns $\{E_R^s,E_R^p,E_{in}^{s,-},E_{in}^{s,+}, E_{in}^{p,-}, E_{in}^{p,+},E_T^s,E_T^p\}$ through the boundary conditions, at the two interfaces $z=0$ and $z=-d$:
\begin{equation}\label{eqn:boundary_electric1}
    \hat{\bm z} \times \left.\left(\bm{E}_I + \bm{E}_R - \bm{E}_{in}\right)\right|_{z=0} = \bm{0}~,
\end{equation}
\begin{equation}\label{eqn:boundary_magnetic1}
    \hat{\bm z} \times \left.\left(\bm{B}_I + \bm{B}_R - \bm{B}_{in}\right)\right|_{z=0} = \frac{4\pi}{c}\bm{J}^{(1)}~,
\end{equation}
\begin{equation}\label{eqn:boundary_electric2}
    \hat{\bm z} \times \left.\left(\bm{E}_{in} - \bm{E}_{T}\right)\right|_{z=-d} = \bm{0}~,
\end{equation}
\begin{equation}\label{eqn:boundary_magnetic2}
    \hat{\bm z} \times \left.\left(\bm{B}_{in} - \bm{B}_{T}\right)\right|_{z=-d} = \frac{4\pi}{c}\bm{J}^{(2)}~.
\end{equation}
with:
\begin{equation}\label{constitutive_app}
    J^{(\ell)}_{i}(\bm k,\omega) = \sigma^{(\ell,\ell^\prime)}_{ij}(\bm k,\omega) E^{(\ell^\prime)}_{j}(\bm k,\omega)~.
\end{equation}

We stress how, due to the constitutive equation~\eqref{constitutive_app}, the boundary conditions~\eqref{eqn:boundary_magnetic1} and \eqref{eqn:boundary_magnetic2} couple the fields in the upper region $\bm{E}_I + \bm{E}_R$ to the field in the bottom region $\bm{E}_T$ in a non-trivial way through the tunneling between the layers as the microscopic mechanism. In what follows we consider a conductivity tensor of the form presented in the main text, Eq.~\eqref{eqn:constitutive1}.

The reflection and transmission tensors can be calculated exactly, however we perform a series expansion on $d/\lambda$, where $\lambda$ is the typical wavelength of the incident light. This approximation is generally justified since $d\approx 0.335~{\rm nm}$. To first order, we obtain the following reflection and transmission tensors elements:
\begin{equation}
    r^{s,s}=r^{s,s}_{\rm SL}-\frac{2 i d k_{z,1}}{\varepsilon_{2}}\frac{c^4 \varepsilon_{2} \left(k_{z,3}^2-k_{z,2}^2\right)+8 \pi  c^2 \left[2 \pi  k_{z,2}^2 \sigma_{xy}^2+k_{z,3} \omega  \varepsilon_{2} (\sigma_{0}+\sigma_{1})\right]+16 \pi ^2 \omega ^2 \varepsilon_{2} (\sigma_{0}+\sigma_{1})^2}{ \left[c^2 (k_{z,1}+k_{z,3})+8 \pi  \omega  (\sigma_{0}+\sigma_{1})\right]^2}~,
\end{equation}
\begin{equation}
    t^{s,s} = t^{s,s}_{\rm SL}+\frac{2 i d k_{z,1}}{\varepsilon_{2}}\frac{ c^4 \varepsilon_{2} (k_{z,2}^2-k_{z,3}^2) +4 \pi  c^2 \left[\omega  \varepsilon_{2} (k_{z,1}-k_{z,3}) (\sigma_{0}+\sigma_{1})-4 \pi  k_{z,2}^2 \sigma_{xy}^2\right]+16 \pi ^2 \omega ^2 \varepsilon_{2} (\sigma_{0}+\sigma_{1})^2}{\left[c^2 (k_{z,1}+k_{z,3})+8 \pi  \omega  (\sigma_{0}+\sigma_{1})\right]^2}~,
\end{equation}
\begin{equation}
    r^{p,s} = -\frac{8 i \pi  d k_{z,1} \sigma_{xy} \sqrt{\varepsilon_{1}}}{c \varepsilon_{2}}\frac{c^2 \left[4 \pi  k_{z,3}k_{z,2}^2 (\sigma_{0}+\sigma_{1})+k_{z,2}^2\omega  \varepsilon_{3}-k_{z,3}^2 \omega  \varepsilon_{2}\right]-4 \pi  k_{z,3} \omega ^2 \varepsilon_{2} (\sigma_{0}+\sigma_{1})}{\left[c^2 (k_{z,1}+k_{z,3})+8 \pi  \omega  (\sigma_{0}+\sigma_{1})\right] \left[8 \pi  k_{z,1} k_{z,3} (\sigma_{0}+\sigma_{1})+k_{z,1} \omega  \varepsilon_{3}+k_{z,3} \omega  \varepsilon_{1}\right]}~,
\end{equation}
\begin{equation}
    t^{p,s} = -\frac{8 i \pi  d k_{z,1} \sigma_{xy} \sqrt{\varepsilon_{3}}}{c \varepsilon_{2}}\frac{c^2 \left[4 \pi  k_{z,1} k_{z,2}^2 (\sigma_{0}+\sigma_{1})+k_{z,1} k_{z,3} \omega  \varepsilon_{2}+k_{z,2}^2 \omega  \varepsilon_{1}\right]+4 \pi  k_{z,1} \omega ^2 \varepsilon_{2} (\sigma_{0}+\sigma_{1})}{\left[c^2 (k_{z,1}+k_{z,3})+8 \pi  \omega  (\sigma_{0}+\sigma_{1})\right] \left[8 \pi  k_{z,1} k_{z,3} (\sigma_{0}+\sigma_{1})+k_{z,1} \omega  \varepsilon_{3}+k_{z,3} \omega  \varepsilon_{1}\right]}~.
\end{equation}
Where for brevity we have introduced the coefficients of the {\it effective} single layer, obtained as the zeroth order term of the expansion:
\begin{equation}
    r^{s,s}_{\rm SL} = \frac{2 c^2 k_{z,1}}{c^2 (k_{z,1}+k_{z,3})+8\pi \omega  (\sigma_{0}+\sigma_{1})}-1~,
\end{equation}
\begin{equation}
    t^{s,s}_{\rm SL} = \frac{2 c^2 k_{z,1}}{c^2 (k_{z,1}+k_{z,3})+8\pi \omega  (\sigma_{0}+\sigma_{1})}~.
\end{equation}
For the $p-$ polarized incident light we perform a similar calculation, leading to the other four terms:
\begin{equation}\label{eqn:TBG_rpp}
    r^{p,p} = r^{p,p}_{\rm SL}+\frac{2 i d k_{z,1} \varepsilon_{1}}{c^2 \varepsilon_{2}}\frac{c^2 \left\{\left[4 \pi  k_{z,2} k_{z,3} (\sigma_{0}+\sigma_{1})+k_{z,2} \omega  \varepsilon_{3}\right]^2-k_{z,3}^2 \omega ^2 \varepsilon_{2}^2\right\}+16 \pi ^2 k_{z,3}^2 \sigma_{xy}^2 \omega ^2 \varepsilon_{2}}{\left[8 \pi  k_{z,1} k_{z,3} (\sigma_{0}+\sigma_{1})+k_{z,1} \omega  \varepsilon_{3}+k_{z,3} \omega  \varepsilon_{1}\right]^2}~,
\end{equation}
\begin{equation}
\begin{split}
    &t^{p,p}=t^{p,p}_{\rm SL}-\frac{2 i d k_{z,1} \sqrt{\varepsilon_{1}} \sqrt{\varepsilon_{3}}}{\varepsilon_{2} \left[8 \pi  c k_{z,1} k_{z,3} (\sigma_{0}+\sigma_{1})+c \omega  (k_{z,1} \varepsilon_{3}+k_{z,3} \varepsilon_{1})\right]^2}\left\{16 \pi ^2 k_{z,1} k_{z,3} \left[\sigma_{xy}^2 \omega ^2 \varepsilon_{2}+c^2  k_{z,2}^2  (\sigma_{0}+\sigma_{1})^2 \right]\right.\\
    &\left.+ 4 \pi  \omega c^2 (\sigma_{0}+\sigma_{1}) \left[k_{z,2}^2 (k_{z,1} \varepsilon_{3}+k_{z,3} \varepsilon_{1})-2 k_{z,1} k_{z,3}^2 \varepsilon_{2}\right]+\omega ^2 c^2 \left[k_{z,1} k_{z,3} \varepsilon_{2} (\varepsilon_{2}-\varepsilon_{3})+k_{z,2}^2 \varepsilon_{1} \varepsilon_{3}-k_{z,3}^2 \varepsilon_{1} \varepsilon_{2}\right]\right\}~,
\end{split}
\end{equation}
\begin{equation}\label{eqn:TBG_rsp}
    r^{s,p} = \frac{8 i \pi  d k_{z,1} \sigma_{xy} \sqrt{\varepsilon_{1}}}{c \varepsilon_{2}}\frac{c^2 \left\{k_{z,2}^2 \left[4 \pi  k_{z,3} (\sigma_{0}+\sigma_{1})+\omega  \varepsilon_{3}\right]-k_{z,3}^2 \omega  \varepsilon_{2}\right\}-4 \pi  k_{z,3} \omega ^2 \varepsilon_{2} (\sigma_{0}+\sigma_{1})}{ \left[c^2 (k_{z,1}+k_{z,3})+8 \pi  \omega  (\sigma_{0}+\sigma_{1})\right] \left[8 \pi  k_{z,1} k_{z,3} (\sigma_{0}+\sigma_{1})+k_{z,1} \omega  \varepsilon_{3}+k_{z,3} \omega  \varepsilon_{1}\right]}~,
\end{equation}
\begin{equation}
    t^{s,p} = \frac{8 i \pi  d k_{z,1} \sigma_{xy} \sqrt{\varepsilon_{1}}}{c \varepsilon_{2}}\frac{c^2 \left\{k_{z,1} k_{z,3} \omega  \varepsilon_{2}+k_{z,2}^2 \left[4 \pi  k_{z,3} (\sigma_{0}+\sigma_{1})+\omega  \varepsilon_{3}\right]\right\}+4 \pi  k_{z,3} \omega ^2 \varepsilon_{2} (\sigma_{0}+\sigma_{1})}{\left[c^2 (k_{z,1}+k_{z,3})+8 \pi  \omega  (\sigma_{0}+\sigma_{1})\right] \left[8 \pi  k_{z,1} k_{z,3} (\sigma_{0}+\sigma_{1})+k_{z,1} \omega  \varepsilon_{3}+k_{z,3} \omega  \varepsilon_{1}\right]}~.
\end{equation}
Again we have introduced the following definitions:
\begin{equation}
    r^{p,p}_{\rm SL} = 1-\frac{2k_{z,3} \omega  \varepsilon_{1}}{8 \pi  k_{z,1} k_{z,3} (\sigma_{0}+\sigma_{1})+k_{z,1} \omega  \varepsilon_{3}+k_{z,3} \omega  \varepsilon_{1}}~,
\end{equation}
\begin{equation}
    t^{p,p}_{\rm SL} = \frac{2 k_{z,1} \omega  \sqrt{\varepsilon_{1}} \sqrt{\varepsilon_{3}}}{8 \pi  k_{z,1} k_{z,3} (\sigma_{0}+\sigma_{1})+k_{z,1} \omega  \varepsilon_{3}+k_{z,3} \omega  \varepsilon_{1}}~.
\end{equation}
\section{Optical conductivity of twisted bilayer graphene}\label{appendixF}
We briefly summarize the single-particle band model we have used to describe twisted bilayer graphene (TBG) and the formalism used in order to compute the conductivity tensor.

\subsection{TBG single-particle continuum model}
Layer, sublattice, spin, and valley are the four discrete degrees of freedom characterizing single-electron states in TBG. We can take into account valley and spin degrees of freedom by a degeneracy factor $g = 4 = g_{\rm v}g_{\rm s}$, where the spin-degeneracy factor $g_{\rm s}=2$ has been introduced earlier. The single-particle Hamiltonian of TBG is written in the layer/sublattice basis $\{|1A\rangle, |1B\rangle, |2A\rangle, |2B\rangle\}$ as~\cite{appendix_novelli, appendix_cavicchi_PRB_2024}:
\begin{equation}\label{eqn:continuumtot}
    \hat{{\cal H}}_0 = \left(\begin{array}{cc}
       \hat{{\cal H}}^{(1)}  & \hat{U} \\
        \hat{U}^\dagger & \hat{{\cal H}}^{(2)}
    \end{array}\right)~.
\end{equation}
The state $|\ell \tau\rangle$ refers to layer $\ell = 1, 2$ and sublattice index $\tau = A, B$, $\hat{\cal H}^{(\ell)}$ is the intra-layer Hamiltonian for layer $\ell$, and the operator $\hat{U}$ describes inter-layer tunneling. For
small twist angles, the moir\'e length scale $\sim a/\theta$ is much larger than the lattice parameter $a$ of single-layer graphene. This allows us to replace 
$\hat{\cal H}^{(\ell)}$ by its $\bm{k}\cdot \bm{p}$ massless Dirac fermion limit. This low-energy expansion is done around one of the single layer valleys, $K^{(\ell)}/K^\prime{}^{(\ell)}$:
\begin{equation}\label{eqn:continuumintra}
    \hat{{\cal H}}^{(\ell)} = v_{\rm D}\left[{\cal R}_\ell(\theta/2)(\hat{\bm{p}} \mp \hbar \bm{K}_{\ell})\right]\cdot(\pm\sigma_x,-\sigma_y)~.
\end{equation}
Here, $(\pm\sigma_x,-\sigma_y)$ is a vector of $2\times2$ Pauli matrices (the $\pm$ sign referring to the $K$ and $K^\prime$ valleys, respectively), $\hat{\bm p}$ is the momentum operator, $v_{\rm D} = \sqrt{3}|t|a/(2\hbar) \sim 1 \times
10^6$m/s is the Fermi velocity of single-layer graphene, $|t| = 2.78~{\rm eV}$ being the usual single-particle nearest-neighbor hopping of single layer graphene. The vector $\bm{K}_\ell$ appearing in Eq.~\eqref{eqn:continuumintra} is the position of single layer graphene’s valley $K^{(\ell)}$ measured from the moir\'{e} BZ center $\Gamma$:
\begin{equation}\label{k_point}
    \bm{K}_{1,2} = \frac{8\pi}{3 a}\sin{\left(\frac{\theta}{2}\right)}\left(-\frac{\sqrt{3}}{2},\pm\frac{1}{2}\right)~.
\end{equation}
The rotation matrix ${\cal R}_\ell(\theta/2)$ appearing in \eqref{eqn:continuumintra} is given by:
\begin{eqnarray}
    {\cal R}_{\ell=1,2}\left(\theta/2\right) &=& \cos(\mp\theta/2) \mathbb{I}_{2\times2} - i\sin(\mp\theta/2)\sigma_y \nonumber\\
    &=&\left(\begin{array}{cc}
        \cos\theta/2 & \pm\sin\theta/2 \\
        \mp\sin\theta/2 & \cos\theta/2
    \end{array}\right)~.
\end{eqnarray}
The convention adopted is such that $\theta_{\ell=1} = -\theta/2$ and $\theta_{\ell=2} = \theta/2$. The longitudinal displacement between the two layers is taken as zero in order to obtain the AB-Bernal stacking configuration for $\theta=0$.

The $\hat{U}$ operator describes inter-layer hopping and is given by:
\begin{eqnarray}\label{eqn:interlayeroperator}
    \hat{U} &=& \left(\begin{array}{cc}
       u_0  & u_1 \\
       u_1  & u_0
    \end{array}\right) + e^{-i\frac{2\pi}{3}+i\bm{G}_1\cdot\hat{\bm{r}}} \left(\begin{array}{cc}
       u_0  & u_1 e^{i\frac{2\pi}{3}}\\
       u_1e^{-i\frac{2\pi}{3}}  & u_0
    \end{array}\right) +\nonumber\\ &+&e^{i\frac{2\pi}{3}+i\bm{G}_2\cdot\hat{\bm{r}}} \left(\begin{array}{cc}
       u_0  & u_1 e^{-i\frac{2\pi}{3}}\\
       u_1e^{i\frac{2\pi}{3}}  & u_0
    \end{array}\right)~,
\end{eqnarray}
where
\begin{equation}
    \bm{G}_{1,2} = \frac{8\pi}{\sqrt{3} a} \sin\left(\frac{\theta}{2}\right)\left(\pm \frac{1}{2},\frac{\sqrt{3}}{2} \right)~,
\end{equation}
and $u_0$ ($u_1$) are the intra-sublattice  (inter-sublattice) hopping parameters. In general $u_0\neq u_1$. The difference between these two parameters can, in fact, take into account the lattice corrugation of TBG samples~\cite{appendix_koshino_prx_2018,appendix_lucignano_prb_2019,appendix_carr,appendix_carr_rev_2020}. In order to obtain the numerical results of the text, we fixed $u_0 = 79.7$ meV and $u_1 = 97.5$ meV as in~\cite{appendix_koshino_prx_2018}.

Within the continuum model described by the single-particle Hamiltonian in Eq.~\eqref{eqn:continuumtot}, we can construct the projector operators onto the $i$-th layer $\hat{\Pi}^{(i)}$ by making explicit their action on the basis $|\ell\tau\rangle$:
\begin{equation}\label{eqn:projector_def}
    \hat{\Pi}^{(i)}|\ell\tau\rangle = |i \tau\rangle~.
\end{equation}
In particular their matrix form is given explicitly by:
\begin{equation}\label{eqn:projector1}
    \hat{\Pi}^{(1)} = \left(\begin{array}{cc}
        \hat{\mathbb{I}}_{2\times2} & 0 \\
        0 & 0
    \end{array}\right)~,
\end{equation}
\begin{equation}\label{eqn:projector2}
    \hat{\Pi}^{(2)} = \left(\begin{array}{cc}
        0 & 0 \\
        0 & \hat{\mathbb{I}}_{2\times2}
    \end{array}\right)~,
\end{equation}
where $\hat{\mathbb{I}}_{2\times2}$ is the identity operator acting on the sublattice index.

We describe now the calculation of the local conductivity tensor introduced in the main text in Eq.~\eqref{eqn:constitutive1}. 
From the Kubo formula~\cite{appendix_GiulianiVignale} it can be shown that in a 2D crystal, i.e.~a Bloch translationally-invariant system in which the single-particle eigenstates are of the Bloch type, the local conductivity $\sigma_{\alpha,\beta}(\omega)$ is given by two contributions, namely the intra-band and the inter-band ones~\cite{appendix_novelli}:
\begin{equation}\label{eqn:conductivity}
    \sigma_{\alpha\beta}(\omega) = \sigma_{\alpha\beta}^{\rm intra}(\omega) + \sigma_{\alpha\beta}^{\rm inter}(\omega)~,
\end{equation}
and their full expressions are:
\begin{equation}\label{eqn:conductivity_intra}
    \sigma_{\alpha\beta}^{\rm intra}(\omega) = \frac{\hbar}{\pi}\frac{i{\cal D}_{\alpha\beta}}{\hbar\omega + i\eta}~,
\end{equation}
where:
\begin{equation}\label{eqn:drude_weight}
    {\cal D}_{\alpha\beta} =
    \frac{\pi e^2}{\hbar^2} \sum_\lambda \int \frac{d^2\bm{k}}{(2\pi)^2}[-f_{\bm k, \lambda}^\prime]\langle \bm k, \lambda|\frac{\hbar}{m^*}\hat{p}_\alpha|\bm k, \lambda\rangle \langle \bm k, \lambda| \frac{\hbar}{m^*}\hat{p}_\beta|\bm k, \lambda\rangle ~,
\end{equation}
is the Drude weight~\cite{appendix_novelli}, and
\begin{equation}\label{eqn:conductivity_inter-band}
\sigma_{\alpha\beta}^{\rm inter}(\omega) = \frac{ie^2}{\hbar} \sum_{\lambda\neq\lambda^\prime} \int \frac{d^2\bm{k}}{(2\pi)^2}\left[-\frac{f_{\bm k, \lambda} - f_{\bm k, \lambda^\prime}}{\epsilon_{\bm k,\lambda}- \epsilon_{\bm k,\lambda^\prime}}\right] \frac{\langle \bm k, \lambda|\frac{\hbar}{m^*}\hat{p}_\alpha|\bm k, \lambda^\prime\rangle \langle \bm k, \lambda^\prime| \frac{\hbar}{m^*}\hat{p}_\beta|\bm k, \lambda\rangle}{\epsilon_{\bm k,\lambda}- \epsilon_{\bm k,\lambda^\prime} + \hbar \omega + i \eta}~.
\end{equation}
Here $f_{\bm k, \lambda}$ is the Fermi-Dirac distribution, $f_{\bm k, \lambda}^\prime$ is its derivative, $\hat{p}_\alpha$ is the $\alpha$ component of the total momentum operator, $\epsilon_{\bm k,\lambda}$ is the eigenvalue associated with the Bloch eigenstate $|\bm k, \lambda\rangle$ and $\eta > 0$ is a small parameter. Making use of the layer projector operators Eq.~\eqref{eqn:projector1} and Eq.~\eqref{eqn:projector2}, we obtain the layer-resolved conductivity for a bilayer system by introducing the layer-projected momentum operator:
\begin{equation}
    \hat{p}_{\alpha}^{(i)} = \hat{\Pi}^{(i)}\hat{p}_{\alpha}\hat{\Pi}^{(i)}~.
\end{equation}
We then resolve the conductivity tensor as follows:
\begin{equation}\label{eqn:conductivity_layer}
    \sigma_{\alpha\beta}^{(i,j)}(\omega) = \left[\sigma_{\alpha\beta}^{(i,j)}(\omega)\right]^{\rm intra} + \left[\sigma_{\alpha\beta}^{(i,j)}(\omega)\right]^{\rm inter} ~,
\end{equation}
with:
\begin{equation}
    \left[\sigma_{\alpha\beta}^{(i,j)}(\omega)\right]^{\rm intra} = \frac{\hbar}{\pi}\frac{i{\cal D}_{\alpha\beta}^{(i,j)}}{\hbar\omega + i\eta}~,
\end{equation}
where:
\begin{equation}
    {\cal D}_{\alpha\beta}^{(i,j)} =
    \frac{\pi e^2}{\hbar^2} \sum_\lambda \int \frac{d^2\bm{k}}{(2\pi)^2}[-f_{\bm k, \lambda}^\prime]\langle \bm k, \lambda|\frac{\hbar}{m^*}\hat{p}_\alpha^{(i)}|\bm k, \lambda\rangle \langle \bm k, \lambda| \frac{\hbar}{m^*}\hat{p}_\beta^{(j)}|\bm k, \lambda\rangle ~,
\end{equation}
and,
\begin{equation}
\left[\sigma_{\alpha\beta}^{(i,j)}(\omega)\right]^{\rm inter} = \frac{ie^2}{\hbar} \sum_{\lambda\neq\lambda^\prime} \int \frac{d^2\bm{k}}{(2\pi)^2}\left[-\frac{f_{\bm k, \lambda} - f_{\bm k, \lambda^\prime}}{\epsilon_{\bm k,\lambda}- \epsilon_{\bm k,\lambda^\prime}}\right] \frac{\langle \bm k, \lambda|\frac{\hbar}{m^*}\hat{p}_\alpha^{(i)}|\bm k, \lambda^\prime\rangle \langle \bm k, \lambda^\prime| \frac{\hbar}{m^*}\hat{p}_\beta^{(j)}|\bm k, \lambda\rangle}{\epsilon_{\bm k,\lambda}- \epsilon_{\bm k,\lambda^\prime} + \hbar \omega + i \eta}~.
\end{equation}
For the sake of clarity, we explicitly express the conductivity tensor elements introduced in the main text in Eq.~\eqref{eqn:constitutive1} starting from the layer-resolved conductivity of Eq.~\eqref{eqn:conductivity_layer}:
\begin{equation}
    \sigma_{0}(\omega) \equiv \sigma_{xx}^{(1, 1)}(\omega)=\sigma_{yy}^{(1, 1)}(\omega)= \sigma_{xx}^{(2, 2)}(\omega)=\sigma_{yy}^{(2, 2)}(\omega)~,
\end{equation}
\begin{equation}
    \sigma_{1}(\omega) \equiv \sigma_{xx}^{(1, 2)}(\omega)=\sigma_{yy}^{(1, 2)}(\omega)= \sigma_{xy}^{(2, 1)}(\omega)=\sigma_{yx}^{(2, 1)}(\omega)~,
\end{equation}
\begin{equation}
    \sigma_{xy}(\omega) \equiv -\sigma_{xy}^{(1, 2)}(\omega)=\sigma_{yx}^{(1, 2)}(\omega)= \sigma_{xy}^{(2, 1)}(\omega)=-\sigma_{yx}^{(2, 1)}(\omega)~.
\end{equation}
\begin{figure}
    \centering
    \begin{tabular}{ll}
    \begin{overpic}[width=0.33\textwidth]{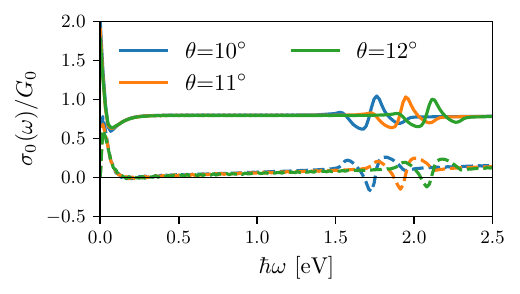}%
    \put(0,54){(a)}%(665,550)
    \end{overpic}
    \begin{overpic}[width=0.33\textwidth]{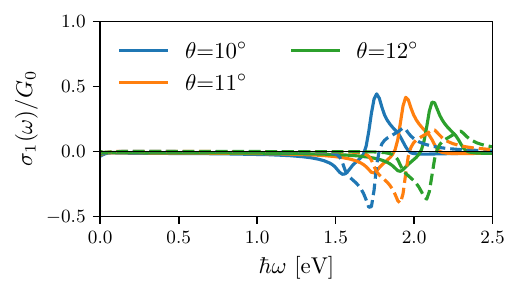}%
    \put(0,54){(b)}%(665,550)
    \end{overpic}
    \begin{overpic}[width=0.33\textwidth]{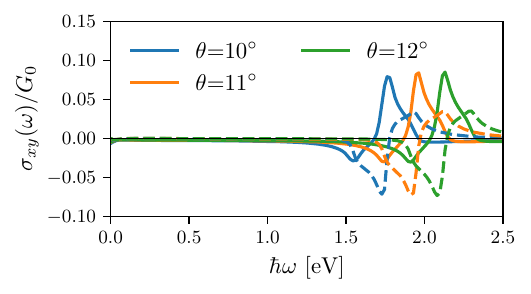}%
    \put(0,54){(c)}%(665,550)
    \end{overpic}
    \end{tabular}
    \caption{Layer resolved optical conductivity entries $\sigma_0(\omega)$ (panel (a)), $\sigma_1(\omega)$ (panel (b)), $\sigma_{xy}(\omega)$ (panel (c)) as a function of the frequency $\omega$ for a few values of the twist angle $\theta$ and in units of $G_0 = 2e^2/h$. These results have been obtained from Eq.~\eqref{eqn:conductivity_layer} for undoped TBG. The temperature is fixed to $T=300~{\rm K}$. In each plot the solid (dashed) line refers to the real (imaginary) part of $\sigma_i(\omega)$, $i=0,1,xy$.}
    \label{fig:SM1}
\end{figure}
Fig.~\ref{fig:SM1} shows numerical results for the optical conductivity entries $\sigma_0$, $\sigma_1$ and $\sigma_{xy}$ obtained within the continuum model presented in this appendix. 

\end{document}